\documentclass[a4paper,10pt]{article}
\pdfoutput=1 % if your are submitting a pdflatex (i.e. if you have
\usepackage{jheppub} % for details on the use of the package, please
\usepackage[T1]{fontenc} % if needed

\usepackage{amsmath,graphicx,verbatim,epsfig}
\usepackage[utf8x]{inputenc}
\usepackage{bm}
\usepackage{cancel}

\newcommand{\Li}{\text{Li}}
\newcommand{\ot}{\leftarrow}

\newcommand{\eps}{\varepsilon}

\newcommand{\nn}{\nonumber}

\newcommand{\bn}{{\bar n}}

\newcommand{\cd}{\cdot}

\newcommand{\be}{\begin{equation}}
\newcommand{\ee}{\end{equation}}
\newcommand{\bea}{\begin{eqnarray}}
\newcommand{\eea}{\end{eqnarray}}
\newcommand{\balign}{\begin{align}}
\newcommand{\ealign}{\end{align}}
\newcommand{\as}{\alpha_s}

\newcommand{\sandwich}[3]{\left< #1 \right | #2 \left | #3 \right >}

\newcommand{\bg}{\begin{gather}}
\newcommand{\foma}{\end{gather}}
\newcommand{\noopsort}[1]{}

\newcommand{\vecb}[1]{\mbox{\boldmath $#1$}}
\newcommand{\vecbe}[1]{\mbox{\boldmath ${\scriptstyle #1}$}}

\def\<{\langle}
\def\>{\rangle}

\def\b{\beta}
  \def\G{\Gamma}
\def\d{\delta}

\def\m{\mu}

\def\t{\tau}

\def\({\left(}
\def\[{\left[}
\def\){\right)}
\def\]{\right]}

\def\ln{\hbox{ln}}

\def \le { \left    }
\def \ri { \right }

\usepackage[normalem]{ulem} % \sout{old text} for strikeout
\renewcommand\sout{\bgroup \color[rgb]{1,0,0} \ULdepth=-.5ex \ULset}

\newcommand{\eq}[1]{Eq.~\eqref{#1}}

\title{\boldmath Unpolarized Transverse Momentum Dependent Parton Distribution  and Fragmentation Functions at next-to-next-to-leading order}

\author[a]{Miguel G. Echevarria,}
\author[b]{Ignazio Scimemi}
\author[c]{and Alexey Vladimirov}

\affiliation[a]{Departament de F\'isica Qu\`antica i Astrof\'isica and
Institut de Ci\`encies del Cosmos, Universitat de Barcelona,\\
Mart\'i i Franqu\`es 1, 08028 Barcelona, Spain}
\affiliation[b]{Departamento de F\' isica Te\'orica II, Universidad Complutense de Madrid,\\
Ciudad Universitaria, 28040 Madrid, Spain}
\affiliation[c]{Institut f\"ur Theoretische Physik, Universit\"at Regensburg,\\
D-93040 Regensburg, Germany}

\emailAdd{mgechevarria@icc.ub.edu}
\emailAdd{ignazios@fis.ucm.es}
\emailAdd{aleksey.vladimirov@gmail.com}

\abstract{

The transverse momentum dependent parton distribution/fragmentation functions (TMDs) are essential in the factorization of a number of processes like Drell-Yan scattering, vector boson production, semi-inclusive deep inelastic scattering, etc.  
We provide a comprehensive study of unpolarized TMDs at next-to-next-to-leading order, which includes an explicit calculation of these TMDs and an extraction of their matching coefficients onto their integrated analogues, for all flavor combinations. 
The obtained matching coefficients are important for any kind of phenomenology involving TMDs. 
In the  present study  each individual TMD is calculated without any reference to a specific process.
We recover the known results for parton distribution functions and provide new results for the fragmentation functions.
The results for the gluon transverse momentum dependent fragmentation functions are presented for the first time at one and two loops. 
We also discuss the structure of singularities of TMD operators and TMD matrix elements, crossing relations between TMD parton distribution functions
and TMD fragmentation functions, and renormalization group equations. 
In addition, we consider the behavior of the matching coefficients at threshold and make a conjecture on their structure to all orders in perturbation theory.}

\begin{document}
\maketitle
\flushbottom

%%%%%%%%%%%%%%%%%%%%%%%%%%%%%%%%
%%%%%%%%%%%%%%%%%%%%%%%%%%%%%%%%
%%%%%%%%%%%%%%%%%%%%%%%%%%%%%%%%
\section{Introduction}
\label{sec:intro}
%%%%%%%%%%%%%%%%%%%%%%%%%%%%%%%%

The transverse momentum dependent parton distribution and fragmentation functions (TMDs) play a central role in our understanding of QCD
dynamics in multi-differential cross sections and spin physics. Recently, factorization theorems for Drell-Yan, Vector Boson/Higgs
Production, Semi-Inclusive Deep Inelastic Scattering (SIDIS) and $e^+e^-\rightarrow 2\; hadrons$ processes, both for spin-dependent and
unpolarized hadrons, has been reformulated in terms of individually well-defined TMDs
\cite{Collins:2011zzd,GarciaEchevarria:2011rb,Echevarria:2012js,Echevarria:2014rua}, updating the pioneering works of Collins and Soper
\cite{Collins:1981uw,Collins:1981uk}.
All these processes are fundamental for current high energy colliders, like the LHC, KEK, SLAC, JLab or
RHIC, and future planned facilities, like the EIC, AFTER@LHC, the LHeC or the ILC.

In this work we focus on unpolarized TMDs, which have received much attention recently, being the simplest functions and for which the relevant
factorization theorems have been explicitly checked at next-to-leading order (NLO), with various quantum numbers, by several groups (see
e.g.~\cite{Echevarria:2014rua,Aybat:2011zv,GarciaEchevarria:2011rb,Vladimirov:2014aja,Bacchetta:2013pqa,Echevarria:2015uaa,Zhu:2013yxa,Becher:2010tm,Chiu:2012ir,Ritzmann:2014mka}). 
The current status at next-to-next-to-leading order (NNLO) investigation for the unpolarized TMDs is more involved. 
Even if previous calculations at this order exist (see e.g.
\cite{Catani:2011kr,Catani:2012qa,Catani:2013tia,Gehrmann:2012ze,Gehrmann:2014yya,Luebbert:2016itl}), no calculation of each \emph{individual} TMD in the sense of~\cite{Collins:2011zzd,GarciaEchevarria:2011rb,Echevarria:2012js,Echevarria:2014rua} at two loops is available. 

So, in this work we provide a comprehensive study of TMDs at NNLO based on a direct calculation of TMD matrix elements at NNLO. 
In particular, our results provide an indirect confirmation of the TMD factorization theorem
and the related structure of rapidity divergences. 
In fact we explicitly confirm that the cancellation of rapidity divergences is realized within one single TMD, and not necessarily in the product of two TMDs, which is important when studying the non-perturbative parts of these quantities.

The TMD factorization theorem at higher orders in perturbative QCD is not trivial. 
In fact, in the calculation one has to deal with several types of divergences (ultra-violet (UV), rapidity and infra-red (IR)), which have to be regularized and disentangled properly.
The TMD factorization theorem offers a strategy to remove the rapidity divergences in order to achieve a well-defined TMDs. 
Recently our group has provided a direct calculation of an individual TMD at NNLO, namely the unpolarized quark transverse momentum dependent fragmentation
function (TMDFF)~\cite{Echevarria:2015usa}, and a complete study of the structure of rapidity divergences at the same order in the soft function~\cite{Echevarria:2015byo}
(see also \cite{Luebbert:2016itl,Li:2016ctv}). 
In this work we complete the calculation of the unpolarized quark TMDFF at NNLO, showing also the details of it and including new results for the gluon TMD fragmentation function.

On the other hand, we also calculate the unpolarized quark and gluon TMD parton distribution functions (TMDPDFs). 
Some properties of the TMDPDFs, like their matching onto integrated parton distribution functions (PDFs) can be found in previous works
\cite{Catani:2011kr,Catani:2012qa,Catani:2013tia,Gehrmann:2012ze,Gehrmann:2014yya}, where they were obtained by decomposing the product of two
TMDs and did not use the fact that each TMD is \emph{per se} calculable. 
In other words, these calculations did not fully exploit the results of the TMD factorization theorem of \cite{Collins:2011zzd,GarciaEchevarria:2011rb,Echevarria:2012js,Echevarria:2014rua}. 
We find a complete agreement between our calculation and the results of \cite{Catani:2011kr,Catani:2012qa,Catani:2013tia,Gehrmann:2012ze,Gehrmann:2014yya},  once the proper combination of collinear and soft matrix elements is considered, which represents a strong check and demonstration of the regulator-independence 
of the matching of each TMD onto its corresponding integrated counterpart.

In this work, we slightly go away from the standard formulation of TMD factorization, which is derived for different processes, towards a universal process-independent definition of TMDs. 
With this aim we introduce the process-independent TMD operators, in analogy with the parton string operators for integrated functions. 
So, the TMDs are the hadron matrix elements of these TMD operators. 
Such a reformulation suggests a new more general look on TMDs, and reveals common points between various approaches.

The formulation of a universal TMD operator is possible due to the process independence of the soft factor, that has been discussed for long time~\cite{Collins:2011zzd,GarciaEchevarria:2011rb,Echevarria:2012js,Echevarria:2014rua,Collins:2004nx} and at NNLO has been explicitly demonstrated in~\cite{Echevarria:2015byo}.
Unlike the usual composite and light-cone operators, the TMD operator is more divergent.
It contains rapidity divergences together with UV divergences.
Therefore, the proper definition of TMD operator should include not only UV renormalization constant, but also a mechanism for removing the rapidity divergences.
It is known that the structure of rapidity divergences within the TMDs is similar to the structure of UV divergences.
The similarity is seen already for the soft factor, the logarithm of which is necessarily linear in rapidity divergences, and hence satisfies a renormalization group equation (RGE) with respect to rapidity scale.
Thus, the rapidity divergences can be removed analogously to UV divergences by the ``rapidity renormalization'' factor, which in fact naturally appears in any formulation of TMD factorization theorem (see e.g. \cite{Chiu:2011qc,Chiu:2012ir}).

The matrix elements of TMD operators are free from operator divergences and can have only standard IR divergences related to the external states, as checked here at NNLO for all possible unpolarized TMD operators.
Using the expressions for partonic TMD matrix elements we can extract the matching coefficients of the TMDs onto their corresponding integrated functions (parton distribution functions (PDFs) or fragmentation functions (FFs)).
These coefficients are free from any type of divergences and have a direct impact on phenomenological analyses.
We provide the two-loop coefficients for all unpolarized TMDs, and this is the main practical result of this paper.

The regularization of rapidity divergences used in this paper is  the same as in~\cite{Echevarria:2015usa, Echevarria:2015byo}.
We use the so-called $\d$-regulator, which in practice is just a shift of the residue of the Wilson lines by an amount $\d$, to be removed at the end of the calculation.
Several technical details are necessary for the proper implementation of this regulator at higher orders, which are discussed in the text.
For the rest of the divergences we use the standard dimensional regularization.
This particular choice of regulators simplifies significantly the calculation.
For example, one can avoid a direct calculation of pure virtual contributions, which reduces the number of diagrams to be computed.
The soft function presented in~\cite{Echevarria:2015byo} is a key element for the NNLO calculation of all (polarized and unpolarized) TMDs.
The present calculation is a confirmation of the universality of the soft function as it enters  at the same footing in the calculation of both TMDPDFs and TMDFFs.

We report on the structure of the matching coefficients of the TMDs onto their corresponding integrated functions, consistently with their
RGEs. Also we  consider a series of technical topics which we think are interesting for the expert reader. So, we discuss the realization of the
Gribov-Lipatov correspondence between TMDPDFs and TMDFFs, due to crossing symmetry. Although the computations for TMDPDFs and TMDFFs have been
done independently, we have used the crossing symmetry at intermediate steps of the calculation as a check of our  results. Using the obtained
results we are also able to  formulate a conjecture about the behavior of the coefficient at threshold at all order in perturbation theory. This
 result can be important to study the  perturbative series  for the coefficients at  N$^3$LO.

The paper is organized as follows. Section~\ref{sec:II} provides the definitions of universal TMD operators, with their renormalization. 
We also discuss the basic structure of the small-$b_T$ operator product expansion (OPE), and its relation to the matching procedure. In
Sec.~\ref{sec:III} we discuss the regularization method, the general structure of rapidity divergences as well as the details of the calculation
of TMDs. In Sec.~\ref{sec:RGE}, the renormalization group equations for TMDs are introduced. In Sec.~\ref{sec:1loop} we present in detail the
NLO computation of the TMDs and their matching coefficients. This section serves mainly as a pedagogical demonstration of all the steps of the
computation and the internal structure of TMDs. The technics used for the NNLO calculation are presented in Sec.~\ref{sec:details}, while in
Sec.~\ref{sec:results} we collect all the expressions for TMD matching coefficients for both TMDPDFs and TMDFFs up to NNLO. We study the
coefficient   for large values of the Bjorken variables $x,\; z$ in Sec.~\ref{sec:conj}. Finally, in Sec.~\ref{sec:Conclusions} we conclude.
The set of appendices includes several necessary definitions, some intermediate expressions and side results which were used in the paper.

%%%%%%%%%%%%%%%%%%%%%%%%%%%%%%%%
%%%%%%%%%%%%%%%%%%%%%%%%%%%%%%%%
%%%%%%%%%%%%%%%%%%%%%%%%%%%%%%%%
\section{TMD operators}
\label{sec:II}
%%%%%%%%%%%%%%%%%%%%%%%%%%%%%%%%

%%%%%%%%%%%%%%%%%%%%%%%%%%%%%%%%
\subsection{Definitions of TMD operators}
%%%%%%%%%%%%%%%%%%%%%%%%%%%%%%%%

The factorization theorems for transverse-momentum-dependent cross sections are usually formulated in terms of TMDs.
In this work however we follow a different strategy, namely, we focus our attention on the TMD operators.
Such a consideration allows us to have a homogeneous notation and reveals the similarities between the distribution and fragmentation functions.
It also allows us to formulate statements in a process-independent way.

We define the bare (unrenormalized and rapidity singular) quark, anti-quark and gluon unpolarized TMDPDF operators as follows:
\begin{align}\nn
O^{bare}_q(x,\vecb b_T)&=\frac{1}{2}\sum_X\int \frac{d\xi^-}{2\pi}e^{-ix p^+\xi^-}\left\{T\[\bar q_i \,\tilde W_n^T\]_{a}\(\frac{\xi}{2}\)
~|X\rangle \gamma^+_{ij}\langle X|~\bar T\[\tilde W_n^{T\dagger}q_j\]_{a}\(-\frac{\xi}{2}\)\right\},
\\ \nn
O^{bare}_{\bar q}(x,\vecb b_T)&=\frac{1}{2}\sum_X\int \frac{d\xi^-}{2\pi}e^{-ix p^+\xi^-}\left\{T\[\tilde W_n^{T\dagger}q_j\]_{a}\(\frac{\xi}{2}\)
~|X\rangle \gamma^+_{ij}\langle X|~\bar T\[\bar q_i \tilde W_n^T \]_{a}\(-\frac{\xi}{2}\)\right\},
\\\label{def_PDF_op}
O^{bare}_g(x,\vecb b_T)&=\frac{1}{xp^+}\sum_X\int \frac{d\xi^-}{2\pi}e^{-ix p^+\xi^-}\left\{T\[F_{+\mu} \,\tilde W_n^T\]_{a}\(\frac{\xi}{2}\)|X\rangle \langle X|
\bar T\[\tilde W_n^{T\dagger}F_{+\mu}\]_{a}\(-\frac{\xi}{2}\)\right\},
\end{align}
where $\xi=\{0^+,\xi^-,\vecb b_T\}$, $n$ and $\bn$ are light-cone vectors ($n^2=\bn^2=0,\; n\cdot\bn=1$).
For a generic vector $v$ we have $v^+=\bar n\cd v$ and $v^-=n\cd v$.
The repeated color indices $a$ ($a=1,\dots,N_c$ for quarks and $a=1,\dots,N_c^2-1$ for gluons) are summed up.
The representations of the color SU(3) generators inside the Wilson lines are the same as the representation of the corresponding partons.
The Wilson lines $\tilde W_n^T(x)$ are rooted at the coordinate $x$ and continue to the light-cone infinity along the vector $n$, where it is connected by a transverse link to the transverse infinity (that is indicated by the superscript $T$).
The precise definition of the Wilson lines is given in Sec.~\ref{sec:III}.

The hadronic matrix elements of the operators defined in Eq.~(\ref{def_PDF_op}) provide the unsubtracted TMDPDFs, as they are defined within the TMD factorization theorems \cite{Collins:2011zzd,GarciaEchevarria:2011rb,GarciaEchevarria:2011rb,Echevarria:2012js}:
\begin{eqnarray}\nn
\Phi_{q\leftarrow N}(x,\vecb b_T)&=&\frac{1}{2}\sum_X\int \frac{d\xi^-}{2\pi}e^{-ix p^+\xi^-} \langle N| \left\{T\[\bar q_i \,\tilde
W_n^T\]_{a}\( \frac{\xi}{2} \) |X\rangle \gamma^+_{ij}\langle X|\bar T\[\tilde W_n^{T\dagger}q_j\]_{a}\(-\frac{\xi}{2} \) \right\} | N\rangle,
\\ \nn
\Phi_{\bar q\leftarrow N}(x,\vecb b_T)&=&\frac{1}{2}\sum_X \int \frac{d\xi^-}{2\pi}e^{-ix p^+\xi^-} \langle N|\left\{T\[\tilde
W_n^{T\dagger}q_j\]_{a}\(\frac{\xi}{2} \) |X\rangle \gamma^+_{ij}\langle X|\bar T\[\bar q_i \tilde W_n^T \]_{a}\(-\frac{\xi}{2} \)\right\} |
N\rangle,\nn
\\ \nn
\Phi_{g\leftarrow N}(x,\vecb b_T)&=&\frac{1}{xp^+}\sum_X \int \frac{d\xi^-}{2\pi}e^{-ix p^+\xi^-}
\\&& \times
\langle N| \left\{T\[F_{+\mu} \,\tilde
W_n^T\]_{a}\(\frac{\xi}{2} \)|X\rangle \langle X| \bar T\[\tilde W_n^{T\dagger}F_{+\mu}\]_{a}\( -\frac{\xi}{2} \)\right\} | N\rangle, \label{def_PDF_opsand}
\end{eqnarray}
where $N$ is a nucleon/hadron.
Here the variable $x$ represents the momentum fraction carried by a parton from the nucleon (it also explains the TMD labelling rule $f\ot N$).
One can see at the operator level that TMDs are like the integrated parton densities, with the only difference that parton fields are additionally separated by the space-like distance $b_T$.
The gauge connection between the parton fields follows the path uniquely dictated by the relevant factorization theorems for the given physical processes.

The definition of the operators for the fragmentation functions follows a similar pattern, with the main difference that they should be calculated on final rather than initial states. Formally, one can write
\begin{eqnarray}\nn
\mathbb{O}^{bare}_q(z,\vecb b_T)&=&\frac{1}{4 z N_c}\sum_X\int \frac{d\xi^-}{2\pi}e^{-ip^+\xi^-/z}
\\\nn&&\qquad\qquad\times\langle 0|T\[\tilde W_n^{T\dagger}q_j\]_{a}\(\frac{\xi}{2}\)
|X,\frac{\delta}{\delta J}\rangle \gamma^+_{ij} \langle X,\frac{\delta}{\delta J}|\bar T\[\bar q_i \,\tilde W_n^T\]_{a}\(-\frac{\xi}{2}\)|0\rangle,
\\ \nn
\mathbb{O}^{bare}_{\bar q}(z,\vecb b_T)&=&\frac{1}{4 z N_c}\sum_X\int \frac{d\xi^-}{2\pi}e^{-ip^+\xi^-/z}
\\ \nn && \qquad\qquad\times\langle 0|T\[\bar q_i \,\tilde W_n^T\]_{a}\(\frac{\xi}{2}\)
|X,\frac{\delta}{\delta J}\rangle \gamma^+_{ij} \langle X,\frac{\delta}{\delta J}|\bar T\[\tilde W_n^{T\dagger}q_j\]_{a}\(-\frac{\xi}{2}\)|0\rangle,
\\\label{def_FF_op}
\mathbb{O}^{bare}_g(z,\vecb b_T)&=& \frac{-1}{2(1-\epsilon)p^+ (N_c^2-1)}\sum_X\int \frac{d\xi^-}{2\pi}e^{-ip^+\xi^-/z}
\\ \nn &&
\qquad\qquad\times  \langle 0|T\[\tilde W_n^{T\dagger}F_{+\mu}\]_{a}\(\frac{\xi}{2}\)
|X,\frac{\delta}{\delta J}\rangle  \langle X,\frac{\delta}{\delta J}|\bar T\[F_{+\mu} \,\tilde W_n^T\]_{a}\(-\frac{\xi}{2}\)|0\rangle,
\end{eqnarray}
where $\d/\d J$ is to be understood as the state generated by the variation of the action with respect to the source $J$, which couples to external hadron fields.
Then the unsubtracted TMDFFs are hadronic matrix elements of these operators:
\begin{eqnarray}\nn
\Delta_{q\rightarrow N}(z,\vecb b_T)&=&\frac{1}{4 z N_c}\sum_X\int \frac{d\xi^-}{2\pi}e^{-ip^+\xi^-/z}
\\\nn&& \qquad\qquad\times\langle 0|T\[\tilde W_n^{T\dagger}q_j\]_{a}\(\frac{\xi}{2}\)
|X,N\rangle \gamma^+_{ij} \langle X,N|\bar T\[\bar q_i \,\tilde W_n^T\]_{a}\(-\frac{\xi}{2}\)|0\rangle,
\\ \nn
\Delta_{\bar q\rightarrow N}(z,\vecb b_T)&=& \frac{1}{4 z N_c}\sum_X \int \frac{d\xi^-}{2\pi}e^{-ip^+\xi^-/z}
\\\nn&& \qquad\qquad\times\langle 0|T\[\bar q_i \,\tilde W_n^T\]_{a}\(\frac{\xi}{2}\)
|X,N\rangle \gamma^+_{ij} \langle X,N|\bar T\[\tilde W_n^{T\dagger}q_j\]_{a}\(-\frac{\xi}{2}\)|0\rangle,
\\ \label{def_FF_opsand}
\Delta_{g\rightarrow N}(z,\vecb b_T)&=&\frac{-1}{2(1-\epsilon)p^+(N_c^2-1)}\sum_X\int \frac{d\xi^-}{2\pi}e^{-ip^+\xi^-/z}
\\ &&\nn
\qquad\quad\times\langle 0|T\[\tilde W_n^{T\dagger}F_{+\mu}\]_{a}\(\frac{\xi}{2}\)
\sum_X|X,N\rangle  \langle X,N|\bar T\[F_{+\mu} \,\tilde W_n^T\]_{a}\(-\frac{\xi}{2}\)|0\rangle,
\end{eqnarray}
where again $N$ is a nucleon/hadron. The variable $z$ represents the momentum fraction of the parton carried into the hadron (it also explains
the TMD labelling rule $f\to N$). The definitions of the quark TMDFFs coincide with the one coming from TMD factorization
\cite{Collins:2011zzd,GarciaEchevarria:2011rb,GarciaEchevarria:2011rb,Echevarria:2012js}. To our  knowledge, the gluon TMDFFs were first considered
in~\cite{Mulders:2000sh}. However here we find more convenient to define the normalization factor of the gluon TMDFF in analogy to the
normalization of the integrated FFs.Here the normalization of TMDFFs counts the number of the physical states of a
given flavor (being $2(1-\epsilon)$  the number of physical gluon polarizations in $d=4-2\epsilon$ dimension). Such normalization allows
the crossing relations discussed below to be fulfilled.

Summarizing the expressions Eq.~(\ref{def_PDF_op}-\ref{def_FF_opsand}), the \textit{bare} TMDs\footnote{The term bare TMD is equal to the
common term unsubtracted TMD. Here we use term bare TMD to emphasize its direct relation to the bare operator.} are the hadronic matrix elements of
the corresponding \textit{bare} TMD operator:
\begin{eqnarray}\label{def:TMD_pdf_unsub}
\Phi_{f\ot N}(x,\vecb{b}_T)&=&\langle N|O^{bare}_f(x,\vecb{b}_T)|N \rangle,
\\\label{def:TMD_ff_unsub}
\Delta_{f\to N}(z,\vecb{b}_T)&=&\langle N|^\dagger\mathbb{O}^{bare}_f(z,\vecb{b}_T)|N \rangle^\dagger,
\end{eqnarray}
where the Hermitian conjugation of the states for TMDFFs indicates that these are final states to be placed inside the operator.

Unlike the usual composite or light-like operators which contain only UV divergences, the TMD operators in addition suffer from rapidity
divergences. The UV divergences in the TMDs are removed by the usual renormalization factors. In order to cancel rapidity divergences one should
consider both the zero-bin subtractions and the soft function. According to Soft Collinear Effective Theory (SCET) terminology the
``zero-bin'' represents the soft overlap contribution, that should be removed from the collinear matrix element in order to avoid double
counting of soft singularities~\cite{Manohar:2006nz}. The combination of the zero-bin subtraction with the soft function has a very particular form, which
is dictated by the factorization theorem, and should be included in the definition of the TMD operators as a single ``rapidity renormalization
factor''. Therefore, we introduce the ``rapidity renormalization factor'' $R$, which completes our definition of the \textit{renormalized} TMD
operator. We have
\begin{eqnarray}\nn
O_{q,\bar q}(x,\vecb b_T,\mu,\zeta)=Z_q(\zeta, \mu)R_q(\zeta,\mu)O^{bare}_{q,\bar q}(x,\vecb b_T),
\\\label{renomalized_ops_PDF}
O_{g}(x,\vecb b_T,\mu,\zeta)=Z_g(\zeta, \mu)R_g(\zeta,\mu)O^{bare}_{g}(x,\vecb b_T),
\\\nn
\mathbb{O}_{q,\bar q}(z,\vecb b_T,\mu,\zeta)=Z_q(\zeta, \mu)R_q(\zeta,\mu)\mathbb{O}^{bare}_{q,\bar q}(z,\vecb b_T),
\\\label{renomalized_ops}
\mathbb{O}_{g}(z,\vecb b_T,\mu,\zeta)=Z_g(\zeta, \mu)R_g(\zeta,\mu)\mathbb{O}_{g}^{bare}(z,\vecb b_T),
\end{eqnarray}
where $Z_q$ and $Z_g$ are the UV renormalization constants for TMD operators.
The scales $\mu$ and $\zeta$ are the scales of UV and rapidity subtractions respectively.
While the UV renormalization factors depend on the UV regularization method and the regularization scale $\mu$, the ``rapidity renormalization factors'' also depend on the rapidity regularization method and the rapidity scale $\zeta$.
Moreover, given that the soft function is process independent (as argued with general arguments in~\cite{Collins:2004nx,Collins:2011zzd,GarciaEchevarria:2011rb,Echevarria:2012js,Echevarria:2014rua} and explicitly checked at NNLO in~\cite{Echevarria:2015byo}), the ``rapidity renormalization factors''  are also process independent.
The details of the definition of the factor $R$ are discussed in Sec.~\ref{sec:III}.

It is crucial to observe that both UV and rapidity renormalization factors, $Z$ and $R$ respectively, are the same for TMDPDF and TMDFF
operators. That is not accidental, but the consequence of the fact that both TMDPDF and TMDFF operators have the same local structure (which
makes equal the factors $Z$) and the same geometry of Wilson lines (which makes equal the factors $R$). This significantly simplifies the
consideration of the operators and makes the whole approach more universal. Moreover, from the equality of renormalization factors follows that
the evolution equations for TMDPDF and TMDFF are the same. The appropriate anomalous dimensions can be extracted from $R$ and $Z$, see details
in Sec.~\ref{sec:RGE}.

The renormalization of rapidity divergences needs some caution, since with some regulators the rapidity divergences can be confused with UV
poles. On top of this, the particular form of zero-bin subtractions included in the factor $R$, is regulator dependent. Thus, in order to avoid
any possible confusions we fix the exact order on how to deal with these singular factors: we first remove all rapidity divergences and perform
the zero-bin subtraction, and afterwards multiply by $Z$'s. Such an order implies that the factor $R$ contains not only rapidity divergences,
but also explicit UV divergences which are also taken into account in the factor $Z$.
These two strategies lead to different intermediate expressions, while the final (UV and rapidity divergences-free) expressions are necessarily
the same.

Now, given all previous considerations, we define the \textit{individual} TMDs as
\begin{eqnarray}\label{def:TMD_pdf}
F_{f\ot N}(x,\vecb{b}_T;\mu,\zeta)&=&\langle N|O_f(x,\vecb{b}_T;\mu,\zeta)|N \rangle,
\\\label{def:TMD_ff}
D_{f\to N}(z,\vecb{b}_T;\mu,\zeta)&=&\langle N|^\dagger\mathbb{O}_f(z,\vecb{b}_T;\mu,\zeta)|N \rangle^\dagger.
\end{eqnarray}
Such a definition implies the following relation between bare and renormalized TMDs:
\begin{eqnarray}\label{def:TMD=unTMD_pdf}
F_{f\ot N}(x,\vecb{b}_T;\mu,\zeta)&=&Z_f(\mu,\zeta)R_f(\mu,\zeta)\Phi_{f\ot N}(x,\vecb{b}_T),
\\\label{def:TMD=unTMD_ff}
D_{f\to N}(z,\vecb{b}_T;\mu,\zeta)&=&Z_f(\mu,\zeta)R_f(\mu,\zeta)\Delta_{f\to N}(x,\vecb{b}_T),
\end{eqnarray}
that follows from the TMD factorization theorem.

Finally we comment that,
had we chosen a different order of recombination of singularities, then we would find separate UV-renormalization factors for the soft factor and the collinear matrix element,
which in turn would depend on the parameters of the rapidity regularization. 
Such a strategy has been recently used in \cite{Luebbert:2016itl}, following the ``Rapidity Renormalization Group'' introduced in \cite{Chiu:2011qc,Chiu:2012ir}, which is built in order to cancel the rapidity divergences through renormalization factors from the beam and soft factors independently. 
In our case, however, the rapidity renormalization factor itself is constructed with the soft function. Thus, although at the end of the day the same rapidity logarithms are resummed, the definitions of TMDPDFs and thus the underlying logic is different. In Ref.~\cite{Becher:2010tm,Gehrmann:2012ze,Gehrmann:2014yya} for TMDPDFs the soft  function is hidden in the product of two TMDs.

%%%%%%%%%%%%%%%%%%%%%%%%%%%%%%%%
\subsection{The operator product expansion (OPE) at small $b_T$}
\label{sec:OPE}
%%%%%%%%%%%%%%%%%%%%%%%%%%%%%%%%

The TMDs, as a non-perturbative objects, are a highly involved functions.
Any information on their behavior is important for phenomenological applications.
Apart from evolution equations, QCD perturbation theory can also supply the small-$b_T$ asymptotic behavior of the TMDs, and give their matching coefficient onto their integrated collinear counterparts (see e.g.~\cite{Collins:2011zzd,GarciaEchevarria:2011rb}).
Such a matching is interesting because one expects it to provide a good description of $x/z$-dependence of TMDs in the whole region of $b_T$, and together with a suitable ansatz for the non-perturbative contribution at large-$b_T$, provides a reasonable phenomenological model.
Also the matching represents a strong check of the theory and in this article we explicitly work it out at NNLO, both for quark/gluon distributions and fragmentation TMDs.

At the operator level the small-$b_T$ matching is a statement on the leading term of the small-$b_T$ Operator Product Expansion (OPE). The
small-$b_T$ OPE is a formal operator relation, that relates operators with both light-like and space-like field separation to  operators with
only light-like field separation. It reads
\begin{eqnarray}\label{OPE_formal}
O(\vecb b_T)&=&\sum_n C_n(\vecb b_T,\mu_b)\otimes O_n(\mu_b),
\end{eqnarray}
where $C_n$ are C-number coefficient functions, the $\mu_b$ is the scale of small-$b_T$ singularities factorization or the OPE matching scale
(for simplicity we omit in Eq.~(\ref{OPE_formal}) other matching scales included in the definitions of each one of the pieces of this equation).
The operators on both sides of Eq.~(\ref{OPE_formal}) are non-local along  the same light-cone direction, but the operators $O_n$ are transversely local
while $O(\vecb b_T)$ is transversely non-local.  The operators $O_n$   are all possible operators with proper quantum numbers and can be
organized for instance according to a power expansion. As an example, for quark parton distributions, the most straightforward expansion
consists in the set of two-point operators (in principal one should also include the multi-point operators in the OPE)
\begin{eqnarray}\label{OPE_operator_example}
O_n\sim \frac{1}{2}\int \frac{d\xi^-}{2\pi}e^{-ix p^+\xi^-}\left\{T\[\bar q \,\tilde W_n^T\]_{i,a}\(\frac{\xi}{2}\)
~\gamma^+_{ij}~\(\overleftrightarrow{\partial_T}B_T\)^n\bar T\[\tilde W_n^{T\dagger}q\]_{j,a}\(-\frac{\xi}{2}\)\right\}|_{\vecb b_T=0},
\end{eqnarray}
where the dimension of the transverse derivatives, $\overleftrightarrow{\partial_T}=\overleftrightarrow{\partial}/\partial\vecb b_T$ (these derivatives acts at light-like infinity, therefore the gauge field can be omitted in non-singular gauges), is compensated by some scale $B_T$.
The matching coefficients behave like
\begin{eqnarray}
C_n(\vecb b_T,\mu_b)\sim \(\frac{b_T}{B_T}\)^nf(\ln(\vecb{b}^2_T \mu^2_b)),
\end{eqnarray}
where $f$ is some function.

The unknown scale $B_T$ represents some characteristic transverse size interaction inside the hadron.
So for $b_T\ll B_T$ it is in practice reasonable to consider only the  leading term of the OPE in Eq.~(\ref{OPE_formal}), which gives the matching of the TMDs onto the integrated functions.
The consideration of higher order terms is an interesting and a completely unexplored part of the TMD approach, which we do not further consider in this work.
Note that the OPE onto the operators of the form in Eq.~(\ref{OPE_operator_example}) may not be the most efficient, see discussion and alternative small-$b_T$ OPE based on Laguerre polynomials in \cite{Vladimirov:2014aja,Vladimirov:2014yaa}.

For the TMDPDFs the leading order small-$b_T$ operator (i.e. the operator for the integrated PDF) is just a TMDPDF operator Eq.~(\ref{def_PDF_op}) at $b_T=0$, i.e.
\begin{eqnarray}
O^{bare}_f(x)=O^{bare}_f(x,\vecb{0}_T),
\end{eqnarray}
while for FF kinematics one has an extra normalization factor
\begin{eqnarray}\label{def:int_ff}
\mathbb{O}^{bare}_f(z)=z^{2-2\epsilon}\mathbb{O}^{bare}_f(z,\vecb{0}_T).
\end{eqnarray}
Notice that in the equations above we have dropped a subindex $0$.
In this way the leading terms of the OPEs at small $b_T$ read
\begin{eqnarray}\nn %\label{def_OPE_PDF}
O_f(x,\vecb b_T;\mu,\zeta)&=&\sum_{f'}C_{f\ot f'}(x,\vecb b_T;\mu,\zeta,\mu_b)\otimes O_{f'}(x,\mu_b)+\mathcal{O}\(\frac{b_T}{B_T}\),
\\\label{def_OPE}
\mathbb{O}_f(z,\vecb b_T;\mu,\zeta)&=&\sum_{f'}\mathbb{C}_{f\to f'}(z,\vecb b_T;\mu,\zeta,\mu_b)\otimes \frac{\mathbb{O}_{f'}(z,\mu_b)}{z^{2-2\epsilon}}+\mathcal{O}\(\frac{b_T}{B_T}\),
\end{eqnarray}
where the symbol $\otimes$ is the Mellin convolution in variable $x$ or $z$ , and $f,\;  f'$ enumerate the various flavors of partons.
The running  on the scales $\mu$, $\mu_b$ and $\zeta$ is independent of the regularization scheme and it is dictated by the renormalization group equations.
Taking the hadron matrix elements of the operators we obtain the small-$b_T$ matching between the TMDs and their corresponding integrated functions,
\begin{eqnarray}\nn
F_{f\ot N}(x,\vecb b_T;\mu,\zeta)&=&\sum_{f'}C_{f\ot f'}(x,\vecb b_T;\mu,\zeta,\mu_b)\otimes f_{f'\ot N}(x,\mu_b)
+\mathcal{O}\(\frac{b_T}{B_T}\),
\\\label{def_TMD_expansion}
D_{f\to N}(z,\vecb b_T;\mu,\zeta)&=&\sum_{f'}\mathbb{C}_{f\to f'}(z,\vecb b_T;\mu,\zeta,\mu_b)\otimes \frac{d_{f'\to
N}(z,\mu_b)}{z^{2-2\epsilon}}+\mathcal{O}\(\frac{b_T}{B_T}\).
\end{eqnarray}
The integrated functions (PDFs and FFs) depend only on the Bjorken variables ($x$ for PDFs and $z$ for FFs) and renormalization scale $\mu$,  while all the dependence on  the transverse coordinate $b_T$ and rapidity scale is contained in the matching coefficient and can be calculated perturbatively.

The definition of the integrated PDFs are
\begin{eqnarray}\nn
f_{q\leftarrow N}(x)&=&\frac{1}{2}\sum_X\int \frac{d\xi^-}{2\pi}e^{-ix p^+\xi^-} \langle N| \left\{T\[\bar q_i \,\tilde
W_n^T\]_{a}\( \frac{\xi^-}{2} \) |X\rangle \gamma^+_{ij}\langle X|\bar T\[\tilde W_n^{T\dagger}q_j\]_{a}\(-\frac{\xi^-}{2} \) \right\} | N\rangle,
\\ \nn
f_{\bar q\leftarrow N}(x)&=&\frac{1}{2}\sum_X \int \frac{d\xi^-}{2\pi}e^{-ix p^+\xi^-} \langle N|\left\{T\[\tilde
W_n^{T\dagger}q_j\]_{a}\(\frac{\xi^-}{2} \) |X\rangle \gamma^+_{ij}\langle X|\bar T\[\bar q_i \tilde W_n^T \]_{a}\(-\frac{\xi^-}{2} \)\right\} |
N\rangle,
\\\label{def_PDF}
f_{g\leftarrow N}(x)&=&\frac{1}{xp^+}\sum_X \int \frac{d\xi^-}{2\pi}e^{-ix p^+\xi^-}
\\\nn&& \qquad\qquad\qquad\qquad\times \langle N| \left\{T\[F_{+\mu} \,\tilde
W_n^T\]_{a}\(\frac{\xi^-}{2} \)|X\rangle \langle X| \bar T\[\tilde W_n^{T\dagger}F_{+\mu}\]_{a}\( -\frac{\xi^-}{2} \)\right\} | N\rangle,
\end{eqnarray}
and similarly, for integrated FFs
\begin{eqnarray}\nn
d_{q\rightarrow N}(z)&=&\frac{z^{1-2\eps}}{4 N_c}\sum_X\int \frac{d\xi^-}{2\pi}e^{-ip^+\xi^-/z}
\\\nn&&\qquad\qquad\times \langle 0|T\[\tilde W_n^{T\dagger}q_j\]_{a}\(\frac{\xi^-}{2}\)
|X,N\rangle \gamma^+_{ij} \langle X,N|\bar T\[\bar q_i \,\tilde W_n^T\]_{a}\(-\frac{\xi^-}{2}\)|0\rangle,
\\ \nn
d_{\bar q\rightarrow N}(z)&=&\frac{z^{1-2\eps}}{4 N_c}\sum_X\int \frac{d\xi^-}{2\pi}e^{-ip^+\xi^-/z}
\\\nn&&\qquad\qquad\times \langle 0|T\[\bar q_i \,\tilde W_n^T\]_{a}\(\frac{\xi^-}{2}\)
|X,N\rangle \gamma^+_{ij} \langle X,N|\bar T\[\tilde W_n^{T\dagger}q_j\]_{a}\(-\frac{\xi^-}{2}\)|0\rangle,
\\\label{def_FF}
d_{g\rightarrow N}(z)&=&\frac{-1}{2(1-\epsilon)p^+(N_c^2-1)} \sum_X\int \frac{d\xi^-}{2\pi}e^{-ip^+\xi^-/z}\\
\nn && \qquad\times \langle 0|T\[\tilde W_n^{T\dagger}F_{+\mu}\]_{a}\(\frac{\xi^-}{2}\)
\sum_X|X,N\rangle  \langle X,N|\bar T\[F_{+\mu} \,\tilde W_n^T\]_{a}\(-\frac{\xi^-}{2}\)|0\rangle.
\end{eqnarray}

In practice, in order to calculate the matching coefficients we calculate both sides of Eq.~(\ref{def_OPE}) on some particular states and solve the system for matching coefficients. Since we are interested only in the leading term of the OPE, i.e. the term without transverse derivatives, it is enough to consider single parton matrix elements, with $p^2=0$.  A study of the matching coefficients for higher-derivative operators can be found in~\cite{Vladimirov:2014aja,Vladimirov:2014yaa}.

%%%%%%%%%%%%%%%%%%%%%%%%%%%%%%%%
%%%%%%%%%%%%%%%%%%%%%%%%%%%%%%%%
%%%%%%%%%%%%%%%%%%%%%%%%%%%%%%%%
\section{Regularization and structure of the divergences}
\label{sec:III}
%%%%%%%%%%%%%%%%%%%%%%%%%%%%%%%%

%%%%%%%%%%%%%%%%%%%%%%%%%%%%%%%%
\subsection{Explicit form of rapidity renormalization factor}
%%%%%%%%%%%%%%%%%%%%%%%%%%%%%%%%

In the previous Section we have defined the factor $R_f$ in a rather abstract way, as a kind of ``rapidity renormalization factor''.
In fact, its explicit form is dictated by the TMD factorization theorem and reads
\begin{eqnarray}
R_f(\zeta,\mu)=\frac{\sqrt{S(\vecb{b}_T)}}{\textbf{Zb}},
\end{eqnarray}
where $S(\vecb{b}_T)$ is the soft function and \textbf{Zb} denotes the zero-bin contribution, or in other words the soft overlap of the collinear and soft sectors which appear in the factorization theorem~\cite{Manohar:2006nz,Collins:2011zzd,GarciaEchevarria:2011rb,Echevarria:2012js,Echevarria:2014rua}.
We now elaborate on this definition.

The soft function is defined as a vacuum expectation value of a certain configuration of Wilson lines, which depends on the process under investigation.
For example, for SIDIS it reads
\begin{eqnarray} \label{eq:SF_def}
\tilde S(\vecb b_{T})
=
\frac{{\rm Tr}_c}{N_c} \sandwich{0}{\, T\le[S_n^{T\dagger} \tilde S_\bn^T \ri](0^+,0^-,\vecb b_T) \bar
T\le[\tilde S^{T\dagger}_\bn S_n^T\ri](0)}{0}\,.
\end{eqnarray}
The Wilson lines are defined as usual
\begin{eqnarray}
 \label{eq:SF_def2}
S_{n}^T &=& T_{n} S_{n}\,,
\quad\quad\quad\quad
\tilde S_{\bn}^T = \tilde T_{n} \tilde S_{\bn}\,,
\\ \nn
S_n (x) &=& P \exp \left[i g \int_{-\infty}^0 ds\, n \cdot A (x+s n)\right]\,,
%\\
%&T_{n} (x) = P \exp \left[i g \int_{-\infty}^0 d\tau\, \vec l_\perp \cdot \vec A_{\perp} (\infty^+,0^-,\vec x_\perp+\vec l_\perp \tau)\right]\,,
%\nn
\\\nn
T_{\bn} (x) &=& P \exp \left[i g \int_{-\infty}^0 d\tau\, \vec l_\perp \cdot \vec A_{\perp} (0^+,\infty^-,\vec x_\perp+\vec l_\perp \tau)\right]\,,
\\\nn
\tilde S_\bn (x) &=& P\exp\le[-ig\int_{0}^{\infty} ds\, \bn \cdot A(x+\bn s) \ri]\,,
\\\nn
\tilde T_{n} (x) &=& P\exp\le[-ig\int_{0}^{\infty} d\t\, \vec l_\perp \cdot \vec A_{\perp}(\infty^+,0^-,\vec x_\perp+\vec l_\perp\t) \ri]\,.
\nn
%\\
%&\tilde T_{\bn} (x) = P\exp\le[-ig\int_{0}^{\infty} d\t\, \vec l_\perp \cdot \vec A_{\perp}(0^+,\infty^-,\vec x_\perp+\vec l_\perp\t) \ri]
%\,,
\end{eqnarray}
The transverse gauge links $T_{n}$ are essential for singular gauges, like the light-cone gauge $n \cdot A=0$ (or $\bn \cdot A=0$), see details in Refs.~\cite{Belitsky:2002sm,Idilbi:2010im,GarciaEchevarria:2011md}.
In covariant gauges the transverse links are needed only to preserve the gauge invariance, but in practice do not add any contribution.
Note that collinear Wilson lines $W_n^T(x)$ used in TMD operators Eq.~(\ref{def_PDF_op}-\ref{def_FF_opsand}) are defined in the same way as soft Wilson lines $S_n^T(x)$.
However, we distinguish them since they behave differently under regularization.

The zero-bin (or overlap region) subtraction is a subtle issue.
In fact, the explicit definition of this subtraction significantly depends on the rapidity regularization used (see e.g. discussion in \cite{Echevarria:2012js}).
Thus, for a given regularization scheme it might be even impossible to define the zero-bin as a well-formed matrix element.
Nonetheless, for any regularization scheme it has a very particular calculable expression.
With a conveniently chosen rapidity regularization, the zero-bin subtractions are related to a particular combination of the soft factors.
Using the modified $\delta$-regularization, which is discussed in detail in the next Section, the zero-bin subtraction is literally equal to the SF: $\textbf{Zb}=S(\vecb{b}_T)$.
We should mention that this is not a trivial statement, and in fact, the modified $\delta$-regularization scheme has been adapted such that this relation holds.
In particular, it implies a different regularized form for collinear Wilson lines $W_{n(\bn)}(x)$ and for soft Wilson lines $S_{n(\bn)}(x)$.

So, concluding, in the modified $\delta$-regularization that is used in this work, the expression for the rapidity renormalization factor is
\begin{eqnarray}\label{reg:R=1/S}
R^f(\zeta,\mu)\bigg|_{\delta\text{-reg.}}=\frac{1}{\sqrt{S(\vecb{b}_T;\zeta)}}.
\end{eqnarray}
The relation Eq.~(\ref{reg:R=1/S}) was first checked explicitly at NNLO in~\cite{Echevarria:2015usa,Echevarria:2015byo}, and also confirmed for
various kinematics in this work. We notice that due to the process independence of soft function \cite{Collins:2011zzd,GarciaEchevarria:2011rb,Echevarria:2012js,Echevarria:2014rua,Collins:2004nx}, the factor $R_f$ is also
process independent. The origin of rapidity scale $\zeta$ is explained in the next section.

Let us also make a connection to the formulation of TMDs by Collins in \cite{Collins:2011zzd}. 
In the JCC approach the rapidity divergences are handled by tilting the Wilson lines off-the-light-cone. 
Then the contribution of the overlapping regions and soft factors can be recombined into individual TMDs by the proper combination of different SFs with a partially removed regulator. 
This combination gives the factor $R^f$ in our notation,
\begin{eqnarray}\label{reg:R=SS/S}
R^f(\zeta,\mu)\bigg|_{JCC}=\sqrt{ \frac{\tilde S(y_n,y_c)}{\tilde S(y_c,y_{\bar n}) \tilde S(y_n,y_{\bar n})}}.
\end{eqnarray}
The following logical steps remain the same as with the $\delta$-regulator.

%%%%%%%%%%%%%%%%%%%%%%%%%%%%%%%%
\subsection{Modified $\d$-regularization scheme}
\label{subsec:deltareg}
%%%%%%%%%%%%%%%%%%%%%%%%%%%%%%%%

The original $\delta$-regularization proposed in~\cite{GarciaEchevarria:2011rb} consists in a simple infinitesimal shift of the $i0$-prescriptions in eikonal propagators.
However, such a rude approach appears to be not sufficient at NNLO for several reasons, e.g., the fact that at this order the zero-bin and the soft function are not equal.
Therefore, in~\cite{Echevarria:2015usa,Echevarria:2015byo} the $\delta$-regularization scheme was conveniently modified to overcome this issue.
The modified $\delta$-regularization is implemented at the operator level, and constructed in such a way that it explicitly preserves the non-Abelian exponentiation and the equality of zero-bin and the SF.
The implementation of the regularization at the operator level grants many benefits in the analysis of the all-order structure of rapidity divergences, and allows to prove such statements as the linearity of the logarithm of the soft function in $\ln \delta$.
The detailed discussion on the properties of the modified $\delta$-regularization can be found in~\cite{Echevarria:2015byo}.
Here we limit ourselves to present the definitions and make the essential comments.

The modified $\delta$-regularization scheme has to be defined at the operator level, and consists in modifying the definition of Wilson lines.
So the soft Wilson lines entering the soft function in Eq.~(\ref{eq:SF_def}) are changed according to
\begin{eqnarray}\label{eq:reg}
\tilde S_{\bar n}(0)&=&P\exp\left[ -i g \int_0^\infty d\sigma A_{+} (\sigma \bar n)\right]
%\nn\\
%&
\rightarrow
%&
P\exp\left[ -i g \int_0^\infty d\sigma
A_{+} (\sigma \bar n)e^{-\d^{+}\sigma}\right]\nn
\,,\\
 S_{n}(0)&=&P\exp\left[ i g \int^0_{-\infty} d\sigma A_{-} (\sigma n)\right]
%\nn\\
\rightarrow
P\exp\left[ i g \int^0_{-\infty} d\sigma
A_{-} (\sigma  n)e^{+\d^{-}\sigma}\right]
\,,
\end{eqnarray}
where $\delta^\pm \to +0$.
 At the level of Feynman diagrams in  momentum space, the modified expressions for the eikonal propagators are written
as (e.g. absorption of gluons by a Wilson line $[\infty^+,0]$)
\begin{eqnarray}
\label{eq:reg_propgators}
%&&
\frac{1}{(k_1^+-i0)(k_2^+-i0)...(k_n^+-i0)}  \to
\frac{1}{(k_1^+-i\delta^+)(k_2^+-2i\delta^+)...(k_n^+-ni\delta^+)} \,,
\end{eqnarray}
where the gluons are ordered from infinity to zero (i.e. $k_n$ is the gluon closest to zero). As a consequence of the rescaling invariance of the Wilson lines (that is now explicitly broken by the parameters $\delta^\pm$), the expressions for diagrams in the soft function depend on a single variable $\delta^+\delta^-$. 
The ordering of poles in the eikonal propagators, \eq{eq:reg_propgators}, is crucial for the perturbative exponentiation with usual properties, such as non-abelian exponentiation theorem for color-factors \cite{Gatheral:1983cz,Frenkel:1984pz} or logarithmical counting \cite{Sterman:1981jc}. As a matter of fact, within the modified $\delta$-regularization, only diagrams with non-Abelian color prefactor (\emph{web} diagrams) arise in the exponent. Therefore, the expression for the soft function can be written in the form
\begin{eqnarray}\label{eq:S=exp(S1+S2)}
\tilde S(b_T)=\exp\[a_s C_F\(S^{[1]}+a_sS^{[2]}+...\)\]\,,
\end{eqnarray}
where $a_s=g^2/(4\pi)^2$ is the strong coupling and $C_F$ is the Casimir of the fundamental representation of gauge group $\(C_F=(N^2_c-1)/N_c\text{~for~}SU(N_c)\)$.

The collinear Wilson lines appearing in the definition of the operators, Eq.~(\ref{def_PDF_op})-(\ref{def_FF_op}), should be regularized in a slightly different way in order to accomplish Eq.~(\ref{reg:R=1/S}). This is achieved by rescaling the $\d$-regulator with the Bjorken variables as
\begin{align}
 W_{n}(0)&=P\exp\left[ i g \int^0_{-\infty} d\sigma A_{-} (\sigma n)\right]
%\nn\\
\rightarrow
P\exp\left[ i g \int^0_{-\infty} d\sigma
A_{-} (\sigma  n)e^{+\d^{-} x \sigma}\right]
\,,
\end{align}
in the case of the  Wilson lines appearing in TMDPDFs,  Eq.~(\ref{def_PDF_op}),  and as
\begin{align}
 W_{n}(0)&=P\exp\left[ i g \int^0_{-\infty} d\sigma A_{-} (\sigma n)\right]
%\nn\\
\rightarrow
P\exp\left[ i g \int^0_{-\infty} d\sigma
A_{-} (\sigma  n)e^{+(\d^{-} /z) \sigma}\right]
\,,
\end{align}
in the case of the  Wilson lines appearing in TMDFFs,  Eq.~(\ref{def_FF_op}). This  rescaling is not necessary at NLO, where the contribution of the soft function is multiplied by $\delta(1-x)$ (see details in Sec.~\ref{sec:1loop}), but it is necessary at NNLO and higher orders.

The $\delta$-regularized Wilson line violates the usual rules of gauge transformations. 
This violation is power-suppressed in $\delta$. 
Therefore, throughout the calculation the $\delta$ should be considered an infinitesimal parameter, in order to avoid potential gauge-violating contributions. 
In most part of the calculation this is straightforward, however, the linearly divergent subgraphs should be carefully considered. 
A detailed discussion of this point, as well as other potential issues, can be found in \cite{Echevarria:2015byo}.

The parameter $\zeta$ that appears in the factor $R_f$ is a scale that arises due to the splitting of the soft function among the two TMDs.
In the calculation of the SF, one ends up with a function that depends on $\ln(\mu^2/(\delta^+\delta^-))$.
However, here the $\ln \delta^+$ and $\ln \delta^-$ represent the rapidity divergences related to different TMDs in the TMD factorization theorem. Therefore, one separates these logarithms introducing an extra scale $\zeta$.
In general one has
(see e.g. \cite{Echevarria:2012js,Echevarria:2015byo} for more details)
\begin{eqnarray}
S\(\vecb{b}_T;\ln\(\frac{\mu^2}{\delta^+\delta^-}\)\)=S^{1/2}\(\vecb{b}_T;\ln\(\frac{\mu^2}{(\delta^+/p^+)^2\zeta_+}\)\)
S^{1/2}\(\vecb{b}_T;\ln\(\frac{\mu^2}{(\delta^-/p^-)^2\zeta_-}\)\)
\end{eqnarray}
where $\zeta_+\zeta_-=(p_+p_-)^2=Q^4$, with $Q^2$ being the relevant hard scale of the considered process.
In the calculation of a single TMD (say the TMD oriented along the vector $n$), this operation can be effectively replaced by the substitution
\begin{eqnarray}\label{effective_zeta}
\delta^-=\delta^+\frac{\zeta}{(p^+)^2}.
\end{eqnarray}
Here and in the following we omit the subscripts $\pm$ for the variable $\zeta$.

%%%%%%%%%%%%%%%%%%%%%%%%%%%%%%%%
\subsection{Calculation of TMDs and their matching coefficients onto integrated functions}
%%%%%%%%%%%%%%%%%%%%%%%%%%%%%%%%

In order to calculate the leading matching coefficients of the OPE, we perform the calculation of TMD distributions on parton targets. Since at NNLO all possible flavor channels arise, we need to consider the following TMDs:
\begin{align}
F_{q\ot {q,\bar q, q'}}(x,\vecb b_T,\mu,\zeta)&=Z_{2}^{-1}(\mu)Z_q(\zeta,\mu)R_q(\zeta,\mu)\Phi_{q\ot q,\bar q,q'}(x,\vecb b_T),\nn\\
F_{q\ot  {g}}(x,\vecb b_T,\mu,\zeta)&=Z_{3}^{-1}(\mu)Z_q(\zeta,\mu)R_q(\zeta,\mu)\Phi_{q\ot  g}(x,\vecb b_T),\nn\\
F_{g\ot q}(x,\vecb b_T,\mu,\zeta)&=Z_{2}^{-1}(\mu)Z_g(\zeta,\mu)R_g(\zeta,\mu)\Phi_{g\ot q }(x,\vecb b_T),\nn\\
F_{g\ot g}(x,\vecb b_T,\mu,\zeta)&=Z_{3}^{-1}(\mu)Z_g(\zeta,\mu)R_g(\zeta,\mu)\Phi_{g\ot g }(x,\vecb b_T),\nn\\
D_{q \to {q,\bar q, q'}}(x,\vecb b_T,\mu,\zeta)&=Z_{2}^{-1}(\mu)Z_q(\zeta,\mu)R_q(\zeta,\mu)\Delta_{q\to q,\bar q,q'}(x,\vecb b_T),\nn\\
D_{q \to {g}}(x,\vecb b_T,\mu,\zeta)&=Z_{3}^{-1}(\mu)Z_q(\zeta,\mu)R_q(\zeta,\mu)\Delta_{q\to g}(x,\vecb b_T),\nn\\
D_{g\to q}(x,\vecb b_T,\mu,\zeta)&=Z_{2}^{-1}(\mu)Z_g(\zeta,\mu)R_g(\zeta,\mu)\Delta_{g\to q }(x,\vecb b_T),
\nn\\
D_{g\to g}(x,\vecb b_T,\mu,\zeta)&=Z_{3}^{-1}(\mu)Z_g(\zeta,\mu)R_g(\zeta,\mu)\Delta_{g\to g }(x,\vecb b_T),
\end{align}
where $Z_{2}$ and $Z_{3}$ are the wave function renormalization constant for quarks and gluons, respectively. During the calculation of the partonic matrix elements it is sufficient to put the momentum of the target parton at $p^2=0$. This condition is realized with $\vecb{p}_T=0$ in
 the momentum of target partons and   restricting the  light-cone momentum  component $p^-=0$. Therefore,  the momentum of the target parton is
$p=[p^+,0,\vecb{0}_T]$.

In the following, we denote by a superscript in square brackets the coefficient of the perturbative expansions at a given order, e.g. for the partonic TMDFF
\begin{eqnarray}
D_{f\to f'}(z,\vecb{b}_T,p;\mu,\zeta)=\sum_{n=0}^\infty a_s^n D^{[n]}_{f\to f'}(z,\vecb{b}_T,p;\mu,\zeta)
\,.
\end{eqnarray}
The LO pertubative expansion of TMDs coincides with the unsubtracted matrix element, e.g. for quark-to-quark TMDFF,
\begin{eqnarray}
D^{[0]}_{q\to q}&=&\Delta^{[0]}_{q\to q}.
\end{eqnarray}
At NLO one finds
\begin{eqnarray}\label{eval:1-loop}
D^{[1]}_{q\to q}&=&\Delta^{[1]}_{q\to q}-\frac{S^{[1]}\Delta^{[0]}_{q\to q}}{2}+\(Z_q^{[1]}-Z_2^{[1]}\)\Delta^{[0]}_{q\to q}.
\end{eqnarray}
The second term cancels the rapidity-divergent part from the unsubtracted expression, such that the TMD is finite when $\delta\to 0$.
The last term cancels the UV divergences. After these subtractions the result remains singular for $\epsilon \to 0$ due to the collinear divergences that are part of the parton integrated FF.
At NNLO the structure is richer
\begin{align}\nn
D^{[2]}_{q\to q}&=\Delta^{[2]}_{q\to q}-\frac{S^{[1]}\Delta^{[1]}_{q\to q}}{2}+\frac{3
S^{[1]}S^{[1]}\Delta^{[0]}_{q\to q}}{8}-\frac{S^{[2]}\Delta^{[0]}_{q\to q}}{2}+
\(Z_D^{[1]}-Z_2^{[1]}\)\(\Delta^{[1]}_{q\to q}-\frac{S^{[1]}\Delta^{[0]}_{q\to q}}{2}\)
\\  &\label{eval:2-loop}
+\(Z_D^{[2]}-Z^{[2]}_2-Z_2^{[1]}Z_D^{[1]}+Z_2^{[1]}Z_2^{[1]}\)\Delta^{[0]}_{q\to q}.
\end{align}
All rapidity divergences arise and are canceled in the first line of Eq.~(\ref{eval:2-loop}), while in the second line we  have just UV renormalization constants. In the case of TMDPDFs the perturbative expansion is the same (with the trivial substitution $\Delta_i\to \Phi_i$).

Finally, we calculate the matching of the TMDs onto their corresponding integrated  functions.
At LO the matching coefficients are trivially
\begin{align}\label{eval:LO}
C^{[0]}_{f\ot f'}=\delta_{ff'}\delta(\bar x),
&&
\mathbb{C}^{[0]}_{f\to f'}=\delta_{ff'}\delta(\bar z)
\,,
\end{align}
where $\bar x=1-x$, and $\bar z=1-z$.
Comparing the matrix element at NLO we obtain
\begin{align}\label{eval:C1}
C^{[1]}_{f\ot f'}=F^{[1]}_{f\ot f'}-f^{[1]}_{f\ot f'},
&&
\mathbb{C}^{[1]}_{f\to f'}=D^{[1]}_{f\to f'}-\frac{d^{[1]}_{f\to f'}}{z^{2-2\epsilon}}.
\end{align}
At NNLO we have
\begin{eqnarray}\nn
C^{[2]}_{f\ot f'}&=&F^{[2]}_{f\ot f'}-\sum_r C^{[1]}_{f\ot r}\otimes f^{[1]}_{r\ot f'} -f^{[2]}_{f\ot f'},
\\\label{eval:C2}
\mathbb{C}^{[2]}_{f\to f'}&=&D^{[2]}_{f\to f'}-\sum_r \mathbb{C}^{[1]}_{f\to r}\otimes \frac{d^{[1]}_{r\to f'}}{z^{2-2\epsilon}} -\frac{d^{[2]}_{f\to f'}}{z^{2-2\epsilon}}.
\end{eqnarray}
Notice the factor $z^{2-2\epsilon}$ in the case of TMDFF, which comes from the operator definition in Eq.~(\ref{def:int_ff}).

The matching procedure in Eqs.~(\ref{eval:C1}-\ref{eval:C2}) ensures the cancellation of the IR divergences in the matching coefficients.
In our regularization scheme these divergences are regularized by dimensional regularization.
That is why it is particularly important to know the $\epsilon$ dependence in Eqs.~(\ref{eval:C1}-\ref{eval:C2}) at all orders in $\epsilon$: one can immediately realize that the linear term in $\epsilon$ of the coefficient $C^{[1]}$ in combination with the single pole of $f^{[1]}$ contributes to the finite part of $C^{[2]}$.
Also, the coefficient $z^{-2\epsilon}$ gives a non-trivial  contribution  when combined  with  the poles of $d^{[1,2]}$.

%%%%%%%%%%%%%%%%%%%%%%%%%%%%%%%%
%%%%%%%%%%%%%%%%%%%%%%%%%%%%%%%%
%%%%%%%%%%%%%%%%%%%%%%%%%%%%%%%%
\section{Renormalization Group Equations}
\label{sec:RGE}
%%%%%%%%%%%%%%%%%%%%%%%%%%%%%%%%

%%%%%%%%%%%%%%%%%%%%%%%%%%%%%%%%
\subsection{Anomalous dimensions of TMD operators}
\label{sec:RGE_AD}
%%%%%%%%%%%%%%%%%%%%%%%%%%%%%%%%

The renormalization group equations (RGEs) fix the scale dependence of the matching coefficients of the TMDs onto integrated functions, and follow from the very definition of the OPE, i.e. Eq.~(\ref{def_TMD_expansion}).
Differentiating both sides of Eq.~(\ref{def_OPE}) with respect to the scales we obtain the RGEs for the matching coefficients in terms of the anomalous dimensions of TMD operators and integrated operators. The anomalous dimension of TMD operators is defined as
\begin{align}
\mu^2 \frac{d}{d\mu^2}O_f(x,\vecb b_T)=\frac{1}{2}\gamma^f(\mu,\zeta)O_f(x,\vecb b_T),
&&
\label{RGE:mu}
\mu^2 \frac{d}{d\mu^2}\mathbb{O}_f(z,\vecb b_T)=\frac{1}{2}\gamma^f(\mu,\zeta)\mathbb{O}_f(z,\vecb b_T).
\end{align}
Both the TMDPDF and TMDFF operators have the same anomalous dimension, as a result of the universality of the hard interactions~\cite{Collins:2011zzd,GarciaEchevarria:2011rb,Echevarria:2012js}. The anomalous dimension $\gamma_f$ comes solely from the renormalization factor $Z_f$. Using the standard RGE technique we obtain
\begin{eqnarray}
\gamma^q(\mu,\zeta)=2\,\widehat{AD}\(Z_2-Z_q\),\qquad \gamma^g(\mu,\zeta)=2\,\widehat{AD}\(Z_3-Z_g\),
\end{eqnarray}
where $\widehat{AD}$ represents the operator which extracts the  anomalous dimension  from the counterterm (i.e. it gives the coefficient of the
first pole in $\epsilon$ with $n!$ prefactor, being $n$ the order of the perturbative expansion). The prefactor $2$ arises from the
normalization of anomalous dimension Eq.~(\ref{RGE:mu}).

The flow with respect to the rapidity parameter follows from the factor $R$, and it is also the same for both types of operators, due to the universality of the soft interactions (see discussion in \cite{Echevarria:2015byo}, also in \cite{Chiu:2011qc}),
\begin{align}
\zeta \frac{d}{d\zeta}O_f(x,\vecb b_T)=-\mathcal{D}^f(\mu,\vecb{b}_T)O_f(x,\vecb b_T),
&&
\label{RGE:zetaO}
\zeta \frac{d}{d\zeta}\mathbb{O}_f(z,\vecb b_T)=-\mathcal{D}^f(\mu,\vecb{b}_T)\mathbb{O}_f(z,\vecb b_T).
\end{align}
The representation independence of  non-Abelian exponentiation implies the so-called Casimir scaling of anomalous dimension $\mathcal{D}$, see \cite{Echevarria:2015byo}:
\begin{eqnarray}
\frac{\mathcal{D}^q}{\mathcal{D}^g}=\frac{C_F}{C_A}=\frac{N_c^2-1}{2N_c^2}.
\end{eqnarray}
It is worth to mention that RGEs for TMD operators, in contrast to RGEs for integrated operators, do not mix the operators of different flavors. The rapidity anomalous dimension $\mathcal{D}^f$ can be extracted solely from the prefactor $R_f$ \cite{Echevarria:2015byo} as
\begin{eqnarray}
\mathcal{D}^f(\mu,\zeta)=-\frac{d \ln R_f}{d \ln\zeta}\Big|_{f.p}=-\frac{1}{2}\frac{d \ln R_f}{d \ln \delta^+}\Big|_{f.p},
\end{eqnarray}
where $f.p.$ denotes the extraction of the finite part, i.e. neglecting the poles in $\epsilon$.
The singular part of the factor $R$ is related to the renormalization factor as follows:
\begin{eqnarray}
\frac{d \ln R_f}{d\ln \zeta}\Big|_{s.p.}=\frac{d \ln Z_f}{d \ln \mu^2},
\end{eqnarray}
where $s.p.$ denotes the extraction of the singular part, i.e. the poles in $\epsilon$.

Note that these relations are independent of the regularization procedure, i.e. they hold for any rapidity regularization scheme. In the modified $\delta$-regularization the explicit expressions for the soft function (and hence for the factor $R_f$) are presented in Appendices~\ref{app:NLO} and \ref{app:NNLO}.
All relations in this Subsection are explicitly checked at NLO and NNLO, and the resulting anomalous dimensions, which are collected in Appendix~\ref{app:ADs}, coincide with the known values.

The consistency of the differential equations (\ref{RGE:mu}-\ref{RGE:zetaO}) implies that the cross-derivatives of the anomalous dimension are equal to each other (\cite{Echevarria:2015byo,Chiu:2011qc}),
\begin{eqnarray}
\mu^2 \frac{d}{d\mu^2}\(-\mathcal{D}^f(\mu^2,\vecb{b}_T)\)=\zeta\frac{d}{d\zeta}\(\frac{\gamma^f(\mu,\zeta)}{2}\)=-\frac{\Gamma_{cusp}^f}{2}.
\label{eq:cusp1}
\end{eqnarray}
The first terms of the perturbative expansion of the cusp anomalous dimension
$\Gamma_{cusp}^f$ can be found in Appendix~\ref{app:ADs}. From Eq.~(\ref{eq:cusp1}) one finds that the anomalous dimension $\gamma$ is
\begin{eqnarray}
\gamma^f=\Gamma_{cusp}^f\mathbf{l}_\zeta-\gamma_V^f,
\end{eqnarray}
where we introduce the notation
\begin{eqnarray}\label{def_logarithms}
\mathbf{L}_X\equiv \ln\(\frac{X^2\vecb{b}^2_T}{4e^{-2\gamma_E}}\),~~~
\mathbf{l}_X\equiv \ln\(\frac{\mu^2}{X}\),~~~
\pmb \lambda_\delta\equiv\ln\(\frac{\delta^+}{p^+}\).
\end{eqnarray}
At the level of the renormalization factors this relation allows one to unambiguously fix the logarithmic part of the factor $R_f$, by means of the relation~\footnote{A similar one can be found in \cite{Chiu:2011qc,Chiu:2012ir}.}
\begin{eqnarray}
\frac{d^2 \ln R_f}{d\ln \mu^2 \, d\ln \zeta}\Bigg|_{f.p}=\widehat{AD}\left[Z_f\(\frac{d \ln R_f}{d\ln \zeta}\)_{s.p}\right]=-\frac{\Gamma^f_{cusp}}{2}.
\end{eqnarray}

%%%%%%%%%%%%%%%%%%%%%%%%%%%%%%%%
\subsection{RGEs for matching coefficients}
%%%%%%%%%%%%%%%%%%%%%%%%%%%%%%%%

The RGEs for the matching coefficients can be obtained by deriving both sides of Eq.~(\ref{def_OPE}). The only extra information which is needed is the evolution of the light-cone operator.
That is given by DGLAP\footnote{DGLAP is an acronym for Dokshitzer, Gribov, Lipatov, Altarelli, Parisi.} equations
\begin{align}
\mu^2 \frac{d}{d\mu^2}O_f(x)=\sum_{f'}P_{f\ot f'}(x)O_{f'}(x),
&&
\mu^2 \frac{d}{d\mu^2}\mathbb{O}_f(z)=\sum_{f'}\mathbb{P}_{f\to f'}(z)\otimes \mathbb{O}_{f'}(z),
\end{align}
where $P$ and $\mathbb{P}$ are the DGLAP kernels for the PDF and FF respectively. The leading-order expressions are collected in Appendix~\ref{app:ADs}, while NLO expression can be found in \cite{Moch:1999eb,Mitov:2006wy}.

Considering the derivative with respect to $\zeta$ we obtain the $\zeta$-scaling for the matching coefficients ($\mu_b=\mu$)
\begin{eqnarray}
\nn
\zeta \frac{d}{d\zeta}C_{f\ot f'}(x,\vecb{b}_T;\mu,\zeta)=-\mathcal{D}^f(\mu,\vecb{b}_T) C_{f\ot f'}(x,\vecb{b}_T;\mu,\zeta),
\\\label{RGE:zeta}
\zeta \frac{d}{d\zeta}\mathbb{C}_{f\to f'}(z,\vecb{b}_T;\mu,\zeta)=-\mathcal{D}^f(\mu,\vecb{b}_T)  \mathbb{C}_{f\to f'}(z,\vecb{b}_T;\mu,\zeta).
\end{eqnarray}
The solutions of these differential equations  are
\begin{eqnarray}\nn
C_{f\ot f'}(x,\vecb{b}_T;\mu,\zeta)&=& \exp\(-\mathcal{D}^f(\mu,\vecb{b}_T)\mathbf{L}_{\sqrt{\zeta}}\)
\hat C_{f\ot f'}(x,\mathbf{L}_\mu)
\\\label{RGE:zeta_dep}
\mathbb{C}_{f\to f'}(x,\vecb{b}_T;\mu,\zeta)&=& \exp\(-\mathcal{D}^f(\mu,\vecb{b}_T)\mathbf{L}_{\sqrt{\zeta}}\)\hat{\mathbb{C}}_{f\to f'}(z,\mathbf{L}_\mu).
\end{eqnarray}
This defines the reduced matching coefficients $\hat C$ and $\hat{\mathbb{C}}$, and their RGEs are
\begin{eqnarray}\nn
\mu^2 \frac{d}{d\mu^2}\hat C_{f\ot f'}(x,\mathbf{L}_\mu)=\sum_r \hat C_{f\ot r}(x,\mathbf{L}_\mu)\otimes K_{r\ot f'}^f(x,\mathbf{L}_\mu),
\\\label{RGE:C_mu}
\mu^2 \frac{d}{d\mu^2}\hat{\mathbb{C}}_{f\to f'}(z,\mathbf{L}_\mu)=\sum_r \hat{ \mathbb{C}}_{f\to r}(z,\mathbf{L}_\mu)\otimes \mathbb{K}^f_{r\to
f'}(z,\mathbf{L}_\mu)
\,,
\end{eqnarray}
where the kernels $K$ and $\mathcal{K}$ are
\begin{eqnarray}\nn
K^f_{r\ot f'}(x,\mathbf{L}_\mu)&=&\frac{\delta_{rf'}}{2}\(\Gamma_{cusp}^f\mathbf{L}_\mu-\gamma_V^f\)-P_{r\ot f'}(x),
\\
\mathbb{K}^f_{r\to f'}(z,\mathbf{L}_\mu)&=&\frac{\delta_{rf'}}{2}\(\Gamma_{cusp}^f\mathbf{L}_\mu-\gamma_V^f\)-\frac{\mathbb{P}_{r\to f'}(z)}{z^2}.
\end{eqnarray}

Using these equations one can find the  expression for the logarithmical part of the matching coefficients at any given order, in terms of the anomalous dimensions and the finite part of the coefficient at one order lower.
It is convenient to introduce the notation for the $n$-th perturbative order:
\begin{align}
\label{eq:Clognotation}
\hat C^{[n]}_{f\ot f'}(x,\mathbf{L}_\mu)=\sum_{k=0}^{2n}C_{f\ot f'}^{(n;k)}(x)\mathbf{L}_\mu^k,
&&
\hat{\mathbb{C}}^{[n]}_{f\to f'}(x,\mathbf{L}_\mu)=\sum_{k=0}^{2n}\mathbb{C}_{f\to f'}^{(n;k)}(z)\mathbf{L}_\mu^k.
\end{align}
The expressions for the anomalous dimensions, the recursive solution of the RGEs and the explicit expressions for the coefficients $C$ and $\mathbb{C}$ are given in Appendix \ref{appendix:D}.
The known anomalous dimensions and DGLAP kernels allow to fix the logarithmic dependent pieces of the coefficients.
As a result only the coefficients $C_{f\ot f'}^{(n;0)}$ and $\mathbb{C}_{f\to f'}^{(n;0)}$ are necessary to reconstruct their full expressions.

%%%%%%%%%%%%%%%%%%%%%%%%%%%%%%%%
%%%%%%%%%%%%%%%%%%%%%%%%%%%%%%%%
%%%%%%%%%%%%%%%%%%%%%%%%%%%%%%%%
\section{NLO computation}
\label{sec:1loop}

%%%%%%%%%%%%%%%%%%%%%%%%%%%%%%%%

%%%%%%%%%%%%%%%%%%%%%%%%%%%%%%%%
\begin{figure}[t]
\centering
\includegraphics[width=0.35\textwidth]{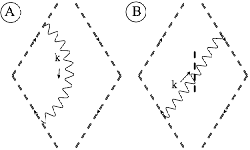}
\caption{Diagrams contributing to soft factor at NLO $S^{[1]}$. The complex conjugated diagrams should be added} \label{fig:SF(1loop)}
\end{figure}
\begin{figure}[t]
\centering
\includegraphics[width=0.55\textwidth]{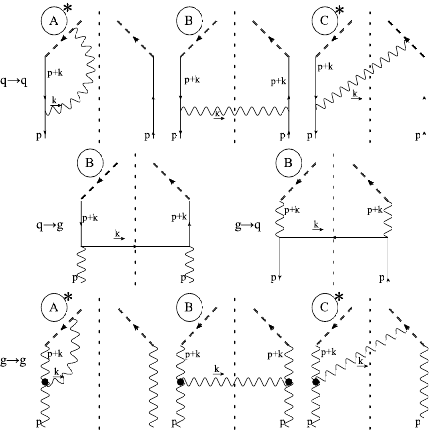}
\caption{All non-zero diagrams contributing to fragmentation function matrix elements at NLO. The external lines represent the final states. The
star on the diagram indicate that complex conjugated diagram should be added. The TMD PDF at NLO is given by the same diagrams with all spinor
arrows pointing opposite direction.} \label{fig:MM(1loop)}
\end{figure}
%%%%%%%%%%%%%%%%%%%%%%%%%%%%%%%%

The calculation of TMDs at NNLO is a complex task. Technically it is convenient and safe to split it in several steps, and perform intermediate
checks. In order to illustrate the procedure and also for pedagogical reasons, in this Section we present the NLO calculation of TMDs and their
matching coefficients, with attention to some important details.
Here and below Feynman gauge is used for the calculations.

The Feynman diagrams for the bare TMDFFs at NLO are drawn in Fig.~\ref{fig:MM(1loop)}. The bare TMDPDFs are given by the same diagrams, but
interpreting the external lines as the initial states and the momentum $p$ as incoming. For the final/initial gluons we choose the polarization
plane perpendicular to $p^\mu$ and $n^\mu$. Thus, the possible diagrams with final/initial gluons radiated by the Wilson lines are zero, and are
not shown in Fig.~\ref{fig:MM(1loop)}.

The only physical Lorentz-invariant scale present in the calculation is $\vecb{b}_T^2$, because the target parton is massless, $p^2=0$, and has no transverse components. The scale $\vecb{b}_T^2$ appears only in the diagrams with  left and right parts connected by gluon/quark exchange.
 Therefore, the pure virtual diagrams (i.e. diagrams without any cut propagator) are zero\footnote{It is not the case for the soft factor, where one has the Lorentz-invariant scale $\delta^+\delta^-$ in addition to $\vecb{b}^2_T$, and thus the virtual diagrams are proportional to $(\delta^+\delta^-)^\epsilon$. However, these contributions completely cancel at all orders in perturbation theory by analogous contributions from diagrams with real quark/gluon exchanges, see proof in \cite{Echevarria:2015byo}.}.
The only piece of the virtual diagrams which is  relevant  for our purposes, is  the  UV-divergent part, that enters the operator renormalization constants $Z_q$ and $Z_g$. The pure virtual diagrams are independent of the kinematics and the operator, which implies that the renormalization constants $Z_q$ and $Z_g$ are the same for PDF and FF operators and independent of $z$ and $x$.
At NLO, the pure virtual diagrams are diagrams $A$ in Fig.~\ref{fig:MM(1loop)}
(for quark-to-quark and gluon-to-gluon sectors), as well as diagram $A$ for the soft factor in Fig.~\ref{fig:SF(1loop)}.
Calculating the ultaviolet limit of the virtual diagrams we obtain
\begin{align}
Z_q^{[1]}=-C_F\(\frac{2}{\epsilon^2}+\frac{4+2 \mathbf{l}_\zeta}{\epsilon}\),
&&
Z_g^{[1]}=-C_A\(\frac{2}{\epsilon^2}+\frac{2+2 \mathbf{l}_\zeta}{\epsilon}\).
\end{align}
Here it is important to preserve the previously defined order of subtraction of  divergences (see Footnote~\ref{foot:2}). So, according to our definition we first recombine the rapidity divergences and then the UV-divergences.

Diagrams $B$ and $C$ in Fig.~\ref{fig:MM(1loop)} provide the quark-to-quark matrix elements,
%\comm{Should it be $\Delta_{f\to f'}(z,\delta)$ and $\Phi_{f\to f'}(z,\delta)$?}
\begin{eqnarray}\nn %\label{1loop:phi}
\Phi_{q\ot q}^{[1]}(x,\d)&=&2C_F \Gamma(-\epsilon)\pmb B^\epsilon \(\bar x(1-\epsilon)+\frac{2x\bar x}{\bar x^2+x^2 \delta^2}\),
\\  \label{eq:T1loop}%\label{1loop:delta}
\Delta_{q\to q}^{[1]}(z,\d)&=&2C_F \Gamma(-\epsilon)\frac{\pmb B^\epsilon}{z^2} \(\bar z(1-\epsilon)+\frac{2z\bar z}{\bar
z^2+\delta^2}\),
\end{eqnarray}
where $\pmb B=\vecb{b}^2_T/4$. We have similar expressions for the  other flavor channels. One can see that  the  expressions  in
Eq.~(\ref{eq:T1loop}) are connected by the relation
\begin{eqnarray}
\Delta^{[1]}_{q\to q}(z,\d)=\frac{-1}{z}\Phi^{[1]}_{q\ot q}\(z^{-1},\d\).
\end{eqnarray}
The validity of this relation to all orders in perturbation theory can be proven in a diagram-by-diagram basis, comparing the expressions in both kinematics. In fact, if we do not remove  any regulator,
the TMDPDFs and the TMDFFs are related to each other by the crossing symmetry $x\leftrightarrow z^{-1}$. This is  the generalization of the well-known Gribov-Lipatov relation between PDF and FF for the TMD operators. We have
%\comm{Should it be $\Delta_{f\to f'}(z,\delta)$ and $\Phi_{f\to f'}(z,\delta)$?}
\begin{eqnarray}\label{def:FF_to_PDF_relation}
\Delta_{f\to f'}(z,\d)=\frac{-1}{z}\mathcal{N}_{f,f'}\Phi_{f\ot f'}\(z^{-1},\d\),
\end{eqnarray}
where the factor  $\mathcal{N}$ arises from the difference of the operator normalization. Comparing the operator definitions we find
\begin{eqnarray}
\mathcal{N}_{q,q}=\mathcal{N}_{g,g}=1,\qquad \mathcal{N}_{q,g}=-(1-\epsilon)\frac{C_F}{T_r},\qquad \mathcal{N}_{g,q}=\frac{-1}{(1-\epsilon)}\frac{T_r}{C_F}.
\end{eqnarray}

Before  combining the  collinear  and soft  matrix element,
we develop  our results  in the limit $\delta\to 0$. This step
allows to  pass from the analytical functions Eq.~(\ref{eq:T1loop}) to the distributions,
where  the singularity at
$z,x\to 1$ is regularized.
Within $\delta$-regularization this step can be done using
\begin{align}\label{1loop:to_plus}
\Phi(x,\delta)=\(\Phi(x,0)\)_++\delta(\bar x)\int_0^1 dy~\Phi(y,\delta)+\mathcal{O}(\delta),
\end{align}
when the functions are regular at $x,\;z\to 0$. In the case that the functions are  singular at $x,z\to 0$ (i.e. TMDFF and gluon distributions) we extract an extra factor of $z$ as
\begin{eqnarray}\label{1loop:to_plus2}
\Delta(z,\delta)=\frac{1}{z}\(z \Delta(z,0)\)_++\delta(\bar z)\int_0^1 dy~y\Delta(y,\delta)+\mathcal{O}(\delta).
\end{eqnarray}
The powers of $\delta$ are irrelevant for our calculation and are dropped.
In the limit $\delta\to 0$ the expressions in Eq.~(\ref{eq:T1loop}) are
%\comm{Should it be $\Delta_{f\to f'}(z,\delta)$ and $\Phi_{f\to f'}(z,\delta)$?}
\begin{eqnarray}\nn
\Phi_{q\ot q}^{[1]}&=&2C_F \Gamma(-\epsilon) \pmb B^{\epsilon}\(\(\frac{2x}{1-x}+\bar x(1-\epsilon)\)_++\delta(\bar
x)\(-\frac{3}{2}-\frac{\epsilon}{2}-2\pmb \lambda_\delta\)\),
\\\label{eq:1loop_after}
\Delta^{[1]}_{q\to q}&=&2C_F \Gamma(-\epsilon)\frac{\pmb B^{\epsilon}}{z^2}\(\(\frac{2z}{1-z}+\bar z(1-\epsilon)\)_++\delta(\bar z)\(-\frac{3}{2}-\frac{\epsilon}{2}-2\pmb \lambda_\delta\)\).
\end{eqnarray}
Let us make a comment on the small-$\delta$ expansion in Eqs.~(\ref{1loop:to_plus})-(\ref{1loop:to_plus2}).
This operation breaks the analytical properties of the calculated functions in the complex plane of $x,\; z$.
Therefore, at this stage of the calculation, one brakes the crossing relation between PDF and FF kinematics in Eq.~(\ref{def:FF_to_PDF_relation}). Indeed, the distributions in Eq.~(\ref{eq:1loop_after}) are not analytical functions of $x$ and $z$ and can not be  analytically continued to each other straightforwardly. That could be done using some regularization method, e.g. by restoring the $\delta$-regularization parameter. This is a simple exercise at NLO but becomes involved at higher orders, see e.g. corresponding analysis for DGLAP kernels in \cite{Stratmann:1996hn}. In practice it results simpler to  calculate the TMDPDFs and the TMDFFs  independently, without using this analytical continuation property.

In order to complete the calculation of the TMDs  we have to include the contribution of the  soft factor, which is computed at NLO and NNLO  in \cite{Echevarria:2015byo}. At NLO the soft function is given by the diagrams shown in Fig.~\ref{fig:SF(1loop)}. The expression for these diagrams, by means of the substitution in Eq.~(\ref{effective_zeta}) is
\begin{eqnarray}
S^{[1]}=-4C_K\pmb B^{\epsilon}\Gamma(-\epsilon)\(\mathbf{L}_{\sqrt{\zeta}}+2\pmb \lambda_\delta-\psi(-\epsilon)-\gamma_E\),
\label{eq:s1L}
\end{eqnarray}
where the color prefactor depends on the representation of Wilson line: $C_K=C_F(C_A)$ when a  quark (gluon) is the initiating  parton.

Combining Eq.~(\ref{eq:1loop_after}) and Eq.~(\ref{eq:s1L}) one can immediately check
 the exact cancellation of the rapidity singularities  in the limit $\delta\to 0$ (represented by $\pmb \lambda_\delta$) between unsubtracted TMDs and the soft factor.

Expanding in $\epsilon$ and combining together all the pieces of  the TMD matrix element according to Eq.~(\ref{eval:1-loop}), we obtain
\begin{eqnarray}\nn
F^{[1]}_{q\ot q}&=&C_F\[\(\frac{-2}{\epsilon}p_{qq}(x)-2\mathbf{L}_\mu p_{qq}(x)+2\bar x\)_++\delta(\bar x)\(-\mathbf{L}_\mu^2+2\mathbf{L}_\mu
\mathbf{l}_\zeta+3\mathbf{L}_\mu+1-\frac{\pi^2}{6}\)+\mathcal{O}(\epsilon)\],
\\\label{eq:1loop}
D^{[1]}_{q\to q}&=&\frac{C_F}{z^2}\Big[\(\frac{-2}{\epsilon}p_{qq}(z)-2\mathbf{L}_\mu p_{qq}(z)+2\bar z\)_+ +\delta(\bar
z)\(-\mathbf{L}_\mu^2+2\mathbf{L}_\mu\mathbf{l}_\zeta+3\mathbf{L}_\mu+1-\frac{\pi^2}{6}\)+\mathcal{O}(\epsilon)\Big],\nn\\
\end{eqnarray}
where $p_{qq}(x)=(1+x^2)/(1-x)$.
This is the final expression for the TMD partonic matrix elements.
They are free from the rapidity and UV divergences, as predicted by the TMD factorization theorem~\cite{Collins:2011zzd,Echevarria:2012js,Echevarria:2014rua}.
The final expressions for unsubtracted TMD at NLO for all other flavor channels are similar and are collected in the Appendix \ref{app:NLO}.

In Eq.~(\ref{eq:1loop}) one recognizes the $\epsilon$-pole, which is part the corresponding integrated functions. In order to complete the
matching between the TMDs and integrated functions we need to calculate the matrix elements of the integrated operators. The diagrams
contributing to these matrix elements are all zero, due to the absence of a Lorentz-invariant scale. Therefore, the only non-zero term is the UV
renormalization factor, which can be deduced from the DGLAP kernel. So, for quark-to-quark channel we have
\begin{eqnarray}
f^{[1]}_{q\ot q}&=&\frac{-2C_F}{\epsilon}\(\frac{1+x^2}{1-x}\)_+,\qquad\qquad  d^{[1]}_{q\to q}=\frac{-2C_F}{\epsilon}\(\frac{1+z^2}{1-z}\)_+.
\end{eqnarray}
The matching prescription of Eq.~(\ref{eval:C1})  allows to derive the coefficients where $\epsilon$-poles are exactly cancelled.
The final matching coefficients for TMDPDFs at LO are
\begin{align}\nn
C^{[0]}_{q\ot q}= \d(1-x), && C^{[0]}_{g\ot g}= \d(1-x),
\end{align}
and the rest are zero.
At NLO we find
\begin{eqnarray}\nn
C^{[1]}_{q\ot q}&=&C_F\[-2\mathbf{L}_\mu p_{qq}(x)+2\bar x+\delta(\bar x)\(-\mathbf{L}_\mu^2+2\mathbf{L}_\mu
\mathbf{l}_\zeta-\frac{\pi^2}{6}\)\],
\\\nn
C^{[1]}_{q\ot g}&=&T_r\(-2\mathbf{L}_\mu p_{gq}(x)+4x\bar x\),
\\\nn
C^{[1]}_{g\ot q}&=&C_F\(-2\mathbf{L}_\mu p_{qg}(x)+2x\),
\\\label{eq:c1Lpdf}
C^{[1]}_{g\ot g}&=&C_A\[-4\mathbf{L}_\mu p_{gg}(x)+\delta(\bar x)\(-\mathbf{L}_\mu^2+2\mathbf{L}_\mu
\mathbf{l}_\zeta-\frac{\pi^2}{6}\)\],
\nn
\\
C^{[1]}_{q\ot q^\prime}&=&0
\,,
\end{eqnarray}
where the definitions of functions $p(x)$ are given in Appendix \ref{app:NLO},  Eq.~(\ref{def_p}).
Hereafter we follow the notation/convention of \cite{Moch:1999eb,Mitov:2006wy}, so that the piece of the coefficients divergent at $x,\; z\to 1$ should be understood as ``plus''-distribution.
The expression for $C^{[1]}_{q\ot q}$ has been already obtained in many articles (see e.g. \cite{Vladimirov:2014aja,Aybat:2011zv,GarciaEchevarria:2011rb,Catani:2011kr,Catani:2012qa,Catani:2013tia,Gehrmann:2012ze,Gehrmann:2014yya,Cherednikov:2008ua,Becher:2010tm,Chiu:2012ir,Ritzmann:2014mka}), the expression for $C^{[1]}_{q\ot g}$ is also well-known \cite{Vladimirov:2014aja,Aybat:2011zv,Catani:2013tia,Gehrmann:2014yya} (note that there is a misprint in \cite{Gehrmann:2014yya}), and the expression for $C^{[1]}_{g\ot q}$ and $C^{[1]}_{g\ot g}$ have been obtained in \cite{Catani:2013tia,Gehrmann:2014yya}.

The matching coefficients for TMDFFs at LO are
\begin{align}\nn
\mathbb{C}^{[0]}_{q\ot q}= \d(1-z), && \mathbb{C}^{[0]}_{g\ot g}= \d(1-z),
\end{align}
 and the rest are zero.
 At NLO we find
\begin{eqnarray}\nn
z^2\mathbb{C}^{[1]}_{q\to q}&=&C_F\[-2p_{qq}(z)\(\mathbf{L}_\mu-2\ln z\)+2\bar z+\delta(\bar
z)\(-\mathbf{L}_\mu^2+2\mathbf{L}_\mu \mathbf{l}_\zeta-\frac{\pi^2}{6}\)\],
\\\nn
z^2\mathbb{C}^{[1]}_{q\to g}&=&C_F\(-2p_{gq}(z)\(\mathbf{L}_\mu-2\ln z\)+2z\),
\\\nn
z^2\mathbb{C}^{[1]}_{g\to q}&=&T_r\(-2p_{qg}(z)\(\mathbf{L}_\mu-2\ln z\)+4z\bar z\),
\\\label{eq:c1Lff}
z^2\mathbb{C}^{[1]}_{g\to g}&=&C_A\[-4(\mathbf{L}_\mu -2\ln z)p_{gg}(z)+\delta(\bar z)\(-\mathbf{L}_\mu^2+2\mathbf{L}_\mu \mathbf{l}_\zeta-\frac{\pi^2}{6}\)\].\nn
\\
z^2\mathbb{C}^{[1]}_{q\to q^\prime}&=&0
\,.
\end{eqnarray}
The functions $p(z)$ are related to the one-loop DGLAP kernels and are defined in Eq.~(\ref{def_p}).
The coefficient $\mathbb{C}_{q\to q}$ has been calculated in \cite{Aybat:2011zv,Echevarria:2014rua}, and $\mathbb{C}_{q\to g}$ agrees with the one calculated in \cite{Aybat:2011zv}.
The coefficients $\mathbb{C}^{[1]}_{g\to q}$ and $\mathbb{C}^{[1]}_{g\to g}$ are presented here for the first time.

One can see that the expression for the matching coefficients for TMDFFs have an extra $\ln(z)$ in comparison to TMDPDFs.
This contribution comes from the difference in the normalization factor $z^{-2\epsilon}$, see Eq.~(\ref{def:int_ff}).
This logarithm is the main source of difference between the TMDFF and TMDPDF matching coefficients.
At higher orders the effects  of the   $z^{-2\epsilon}$ normalization factor are more involved.

%%%%%%%%%%%%%%%%%%%%%%%%%%%%%%%%
%%%%%%%%%%%%%%%%%%%%%%%%%%%%%%%%
%%%%%%%%%%%%%%%%%%%%%%%%%%%%%%%%
\section{NNLO computation}
\label{sec:details}
%%%%%%%%%%%%%%%%%%%%%%%%%%%%%%%%

The  diagrams that contribute to the unsubtracted TMD matrix elements can be generically classified in pure-virtual diagrams (i.e. diagrams with
no cut propagator), virtual-real diagrams (i.e. diagrams with one single cut propagator) and double-real diagrams (i.e. diagrams with two single
cut propagators). Alike NLO case, pure-virtual diagrams are zero due to absence of a Lorentz-invariant scale. Virtual-real and double-real
diagrams are proportional to $\pmb B^{2\epsilon}$. In total, there are about 50 virtual-real diagrams and about 90 double-real diagrams.

The generic expression for virtual-real diagrams is (for TMDFF kinematics)
\begin{eqnarray}
\text{diag}_{VR}=\int \frac{d^dk d^dl}{(2\pi)^{2d}}\frac{\delta\(\frac{\bar
z}{z}p^+-k^+\)e^{i(\vecbe{k}\vecbe{b})_T}\text{Disc}D(k)~~f(k,l,p)~F(n,\delta/z)}{
[(l+p)^2]^{a_1}[(k+p)^2]^{a_2}[(k+l+p)^2]^{a_3}[(k+l)^2]^{a_4}[l^2]^{a_5}},
\end{eqnarray}
where for brevity we drop the $i0$-prescription of propagators. The function $F$ contains all ``plus''-components of the momenta and the
parameter $\delta$, while the function $f$ contains only scalar products of momenta. The discontinuity of the propagator is
\begin{eqnarray}
\text{Disc}D(k)=(2\pi)\delta(k^2)\theta(k^-).
\end{eqnarray}
The generic form of a double-real diagram in the same notation takes the form
\begin{eqnarray}
\text{diag}_{RR}=\int \frac{d^dk d^dl}{(2\pi)^{2d}}\frac{\delta\(\frac{\bar
z}{z}p^+-k^+-l^+\)e^{i(\vecbe{k}\vecbe{b})_T}e^{i(\vecbe{l}\vecbe{b})_T}\text{Disc}D(k)~\text{Disc}D(l)~f(k,l,p)~F(n,\delta/z)}{
[(l+p)^2]^{a_1}[(k+p)^2]^{a_2}[(k+l+p)^2]^{a_3}[(k+l)^2]^{a_4}}.
\end{eqnarray}
The functions $f$ can be re-expressed via the propagators, and so the diagrams can be split into several integrals with $\sum_{i}a_i=3$ for virtual-real
diagrams and $\sum_{i}a_i=2$ for double-real diagrams.
In order to decouple the functions $F$ from the scalar loop integrals we introduce the auxiliary unity factor
\begin{eqnarray}
1=\int_{-\infty}^\infty d\omega~ p^+\delta(\omega p^+-l^+).
\end{eqnarray}
With the help of this trick the dependance of functions $F$ on $k^+$ and $l^+$ can be re-written as a function of $z$ and $\omega$, and all  numerators simplify.
The integration over the loop-momenta is straightforward and all non-zero integrals appearing in the calculation are presented in Appendix~\ref{appendix:C}.

In this way, we are left with a set of one-dimensional integrals over $\omega$.
The evaluation of these integrals is technically the most difficult part of the calculation.
Most part of these integrals are evaluated in terms of $\Gamma$-functions and their derivatives, while several are expressed through hypergeometric functions (and one integral in $g\to g$ and $g\to q$ channels that has been expressed via Appell function $F_1$).
All diagrams are calculated in $d=4-2\epsilon$ dimensions.

During the evaluation of the integrals we have used that we need only their asymptotic behavior at $\delta \to 0$.
In order to find the small-$\delta$ limit we expand the eikonal propagators in Mellin-Barnes contour integral around $\delta=0$.
Then we calculate the integrals over $\omega$, and close the contour over the closest to zero poles.
If an integral has a singularity at $z\to 1$ it should be regularized by means of
a ``plus''-distribution (see Eqs.~(\ref{1loop:to_plus}-\ref{1loop:to_plus2}).
The final expression for a diagram takes the generic form
\begin{eqnarray}\label{eval:gen_diag}
\text{diag.}=\pmb
B^{2\epsilon}\(f_1(z,\epsilon)+\(\frac{\delta^+}{p^+}\)^{\epsilon}f_2(z,\epsilon)+\(\frac{\delta^+}{p^+}\)^{-\epsilon}f_3(z,\epsilon)+\pmb
\lambda_\delta f_4(z,\epsilon)+\pmb \lambda^2_\delta f_5(z,\epsilon)\).
\end{eqnarray}
It is important to mention that the functions $f_2$ and $f_3$ exactly cancel in the sum of all diagrams, which we have checked explicitly.
The unsubtracted TMDPDFs can be calculated in the same manner.

Due to the symmetry of the operators, the expressions for TMDPDFs and TMDFFs satisfy the crossing relation Eq.~(\ref{def:FF_to_PDF_relation}) in
a diagram-by-diagram basis. However, since we consider only the leading contribution at $\delta\to 0$, the diagram-by-diagram crossing is
violated, due to the fact that IR singularities and rapidity singularities get different phases during the procedure of analytical continuation.
In the sum of diagrams all terms $f_{2,3}$ cancel, and one can check the crossing relation in Eq.~(\ref{def:FF_to_PDF_relation}) without any
special tricks. Since we calculated TMDPDFs and TMDFFs independently, such a relation grants a very strong check for our results (however we
have not compared the $\delta$-contribution for $q\to q$ and $g\to g$ channels, for the reasons explained earlier).

Having the expressions for the unsubtracted TMDs, we multiply them by the rapidity and UV renormalization factors.
At NNLO this procedure is given by Eq.~(\ref{eval:2-loop}).
The expressions for the factor $Z^{[1]}$ and soft factor $S^{[1]}$ are given in previous Section.
The NNLO expression for the soft factor has been obtained in \cite{Echevarria:2015byo} and is given in Eq.~(\ref{app:S[2]}).
At this stage we also perform the expansion in $\epsilon$ of the expressions.
To perform the renormalization procedure in Eq.~(\ref{eval:2-loop}) we have to calculate the operator renormalization constants at NNLO $Z_q^{[2]}$ and $Z_q^{[2]}$.
They are given by the UV-part of pure-virtual diagrams.
In our calculation we have not calculated these constants explicitly, but found them by demanding the cancellation of UV poles.
We obtain the following expressions:
\begin{eqnarray} \nn
Z_q^{[2]}&=&\frac{2C_F^2}{\epsilon^4}+\frac{C_F}{2\epsilon^3}\(8C_F(2+\mathbf{l}_\zeta)+11C_A-4 T_rN_f\)
+\frac{C_F}{\epsilon^2}\Bigg[2C_F(4+4\mathbf{l}_\zeta+\mathbf{l}_\zeta^2)+
\\\nn&&C_A\(\frac{25}{9}+\frac{\pi^2}{6}+\frac{11}{3}\mathbf{l}_\zeta\)-T_rN_f\(\frac{8}{9}+\frac{4}{3}\mathbf{l}_\zeta\)\Bigg]+\frac{C_F}{\epsilon}\Bigg[
C_F\(\pi^2-12\zeta_3\)+
\\&& C_A\(-\frac{355}{27}-\frac{11\pi^2}{12}+13 \zeta_3+\(-\frac{67}{9}+\frac{\pi^2}{3}\)\mathbf{l}_\zeta\)+T_rN_f\(\frac{92}{27}+\frac{\pi^2}{3}+\frac{20}{9}\mathbf{l}_\zeta\)\Bigg],
\end{eqnarray}
\begin{eqnarray}\nn
Z_g^{[2]}&=&\frac{2C_A^2}{\epsilon^4}+\frac{C_A}{2\epsilon^3}\(C_A(19+8\mathbf{l}_\zeta)-4T_rN_f\)+\frac{C_A}{\epsilon^2}\Bigg[
C_A\(\frac{55}{36}+\frac{\pi^2}{6}+\frac{23}{3}\mathbf{l}_\zeta+2\mathbf{l}^2_\zeta\)+
\\&&\nn T_rN_f\(\frac{1}{9}-\frac{4}{3}\mathbf{l}_\zeta\)\Bigg]+\frac{C_A}{\epsilon}\Bigg[C_A\(-\frac{2147}{216}+\frac{11\pi^2}{36}+\zeta_3+\(-\frac{67}{9}+\frac{\pi^2}{3}\)\mathbf{l}_\zeta\)+
\\&& T_rN_f\(\frac{121}{54}-\frac{\pi^2}{9}+\frac{20}{9}\mathbf{l}_\zeta\)\Bigg]~.
\end{eqnarray}
As was discussed in Sec.~\ref{sec:RGE_AD}, most part of the UV counterterm should be related to the factor $R$ and to the known anomalous dimensions.

Finally, we perform the matching procedure as in Eq.~(\ref{eval:C2}).
The integrated matrix elements are zero due to the absence of a Lorentz-invariant scale and are given solely by their UV renormalization counterterm.
They can be deduced from the DGLAP kernels, and given by
\begin{eqnarray}
f^{[2]}_{f\ot f'}&=&\frac{1}{2\epsilon^2}\(\sum_r P^{(1)}_{f\ot r}\otimes P^{(1)}_{r\ot f'}+\beta^{(1)}P^{(1)}_{f\ot f'}\)-\frac{P^{(2)}_{f\ot f'}}{2\epsilon},
\\
d^{[2]}_{f\to f'}&=&\frac{1}{2\epsilon^2}\(\sum_r \mathbb{P}^{(1)}_{f\to r}\otimes \mathbb{P}^{(1)}_{r\to f'}+\beta^{(1)}\mathbb{P}^{(1)}_{f\to f'}\)-\frac{\mathbb{P}^{(2)}_{f\to f'}}{2\epsilon}.
\end{eqnarray}
The obtained matching coefficients are free from any kind of divergences.
The results of the calculation are presented in next Section.

%%%%%%%%%%%%%%%%%%%%%%%%%%%%%%%%
%%%%%%%%%%%%%%%%%%%%%%%%%%%%%%%%
%%%%%%%%%%%%%%%%%%%%%%%%%%%%%%%%
\section{Expressions for matching coefficients}
\label{sec:results}
%%%%%%%%%%%%%%%%%%%%%%%%%%%%%%%%

In this section we present the expressions for the finite part of the small-$b_T$ matching coefficients. The logarithmic part can be restored by
using the RGEs and is explicitly given in Appendix~\ref{app:RGE_rec}. For completeness we present LO, NLO and NNLO finite parts together.

%%%%%%%%%%%%%%%%%%%%%%%%%%%%%%%%
\subsection{TMD parton distribution functions}
%%%%%%%%%%%%%%%%%%%%%%%%%%%%%%%%

The LO matching coefficients are
\begin{eqnarray}
C^{(0,0)}_{q\ot q}(x)=C^{(0,0)}_{g\ot g}(x)=\delta(1-x),
\end{eqnarray}
and all other flavor configurations are zero at leading order.

The NLO matching coefficients are
\begin{eqnarray}\label{result:NLO_pdf_start}
\nn C^{(1,0)}_{q\ot q}(x)&=&C_F\(2\bar x-\delta(\bar x)\frac{\pi^2}{6}\),\\
\nn C^{(1,0)}_{q\ot g}(x)&=&4T_r x\bar x,\\
\nn C^{(1,0)}_{g\ot q}(x)&=&2C_F x\\
\nn C^{(1,0)}_{g\ot g}(x)&=&-C_A\delta(\bar x)\frac{\pi^2}{6},\\
C^{(1,0)}_{q\ot q'}(x)&=&C^{(1,0)}_{q\ot \bar q}(x)=0 \label{result:NLO_pdf_end}.
\end{eqnarray}
Here, the coefficient $C^{(1,0)}_{q\ot q'}(x)$ is the coefficient with all possible mixing flavor channels. Thus $q'$ can be any quark or
anti-quark, even of the same flavor as $q$. In other words, the matching coefficient for, say, $u\ot u$ is given by the sum $C_{q\ot q}+C_{q\ot
q'}$, as well as the matching coefficient for $u\ot \bar u$ is given by $C_{q\ot \bar q}+C_{q\ot q'}$.

The NNLO matching coefficients are
\begin{align}\nn
C^{(2,0)}_{q\ot q}(x)&=C_F^2\Bigg\{p_{qq}(x)\Bigg[ -20 \Li_3(x)+4\Li_3(\bar x)-12\ln x\Li_2(\bar x)-4\ln \bar x\Li_2(\bar x)-10 \ln^2 x\ln\bar
x
\\ \nn & + 2\ln^2 \bar x \ln x+\frac{3}{2}\ln^2 x+(8+2\pi^2) \ln x+20 \zeta_3\Bigg]+8\bar x\Li_2(\bar x)+\frac{1+x}{3}\ln^3 x -4\bar x \ln x\,\ln \bar x
\\\nn &+  \frac{7x+3}{2}\ln^2 x-2x\ln\bar x+2(1-12 x)\ln x-\bar x\(22+\frac{\pi^2}{3}\)+\frac{\pi^4}{72}\delta(\bar x)
\Bigg\}
\\\nn & +C_FC_A\Bigg\{p_{qq}(x)\Bigg[
8\Li_3(x)-4\Li_3(\bar x)+4 \ln\bar x\Li_2(\bar x)-4\ln x \Li_2(x)-\frac{\ln^3 x}{3}
\\\nn & - \frac{11}{6}\ln^2 x-\frac{76}{9}\ln x+6 \zeta_3-\frac{404}{27}\Bigg]-4\bar x \Li_2(\bar x)-2x\ln^2x+2x \ln \bar x
\\\nn& +(10x+2)\ln x+\frac{44-\pi^2}{3}\bar x+\delta(\bar x)\(\frac{1214}{81}-\frac{67 \pi^2}{36}-\frac{77}{9}\zeta_3+\frac{\pi^4}{18}\)
\Bigg\}
\\ &
+C_FT_rN_f\Bigg\{p_{qq}(x)\[\frac{2}{3}\ln^2 x+\frac{20}{9}\ln x+\frac{112}{27}\]-\frac{4}{3}\bar x+\delta(\bar
x)\(-\frac{328}{81}+\frac{5\pi^2}{9}+\frac{28}{9}\zeta_3\) \Bigg\} \,, \label{result:NNLO_pdf_start}
\end{align}

\begin{align}\nn
\nn C^{(2,0)}_{q\ot g}(x)&=C_A T_r\Bigg\{ p_{qg}(x)\Bigg[4\Li_3(\bar x)-8\Li_3(x)-4 \ln \bar x\Li_2 (\bar x)-4 \ln x\Li_2 (\bar x)-4 \ln \bar
x\ln^2 x+\frac{2}{3}\ln^3 \bar x-6 \zeta_3\Bigg]
\\\nn&+p_{qg}(-x)\Bigg[4\Li_3\(\frac{1}{1+x}\)-4\Li_3\(\frac{x}{1+x}\)+2\Li_3(x^2)-2\ln x \Li_2(x^2)+2\ln^2x\ln(1+x)
\\\nn&-2\ln x\ln^2(1+x)+\frac{2\pi^2}{3}\ln x-2\zeta_3\Bigg]-4x(1+x)\Li_2(1-x^2)-\frac{8(2-3x+9x^2-14 x^3)}{3x}\Li_2(\bar x)
\\\nn&-4x\bar x\(\ln^2\bar x-\frac{\pi^2}{6}\)+\frac{2}{3}(1+2x)\ln^3 x-\(1-4x+\frac{44}{3}x^2\)\ln^2 x+2x(3-4x)\ln\bar x
\\\nn&+\frac{8}{3}\(\frac{7}{2}-5 x-\pi^2 x+\frac{34}{3}x^2\)\ln x+\frac{344}{27x}-\frac{70}{3}+\frac{86}{3}x-\frac{596}{27}x^2
\Bigg\}
\\
&\nn +C_FT_r\Bigg\{p_{qg}(x)\Bigg[-4 \Li_3(\bar x)-4 \Li_3(x)+2\(\ln \bar x-\ln x\)\(\Li_2(\bar x)-\Li_2(x)+\frac{\pi^2}{6}\)-\frac{2}{3}\ln^3 \bar x
\\&+32 \zeta_3\Bigg]+x\bar x\Bigg[4\ln^2 \bar x-8\ln x\ln\bar x-\frac{4\pi^2}{3}\ln x-2\pi^2\Bigg]-\frac{1-2x+4x^2}{3}\ln^3 x
\\\nn&+\(\frac{1}{2}+6x-4 x^2\)\ln^2 x
+\(8+\frac{2\pi^2}{3}+15 x-8x^2\)\ln x-2x(3-4x)\ln\bar x-13+75 x-72 x^2
\Bigg\},
\end{align}

\begin{align}
\nn C^{(2,0)}_{g\ot q}(x)&=C_F C_A\Bigg\{p_{gq}(x)\Bigg[-12\Li_3(x)+8\ln x\Li_2(x)-\frac{2}{3}\ln^3\bar x+2\ln^2 x\,\ln\bar x+2\ln x\,\ln^2\bar x-\frac{11}{3}\ln^2 \bar x
\\\nn &
+26\zeta_3 \Bigg]+ p_{gq}(-x)\Bigg[4\Li_3\(\frac{1}{1+x}\)-4\Li_3\(\frac{x}{1+x}\)+2\Li_3(x^2)-2\ln x\Li_2(x^2)+\frac{152}{9}\ln\bar x
\\&\nn +2\ln^2 x\,\ln(1+x)-2\ln x\,\ln^2(1+x)
-2\zeta_3\Bigg]-2x\Li_2(1-x^2)+\frac{4(22-24+9x-4x^2)}{3x}\Li_2(\bar x)
\\ &\nn-\frac{2(2+x)}{3}\ln^3x-4 x\ln x\ln\bar x+\(12+3x+\frac{8x^2}{3}\)\ln^2x+2x\ln^2 \bar x+\frac{608+66x}{9}\ln\bar x
\\ &\nn-\frac{498-12 x+176 x^2}{9}\ln x-\frac{4(790-791x+268x^2-152x^3)}{27x}-\frac{2\pi^2}{3} x\Bigg\}
\\ &\nn +C_F^2\Bigg\{p_{gq}(x)\Big[\frac{2}{3}\ln^3\bar x+3\ln^2\bar x+16\ln\bar x\Big]-2x\ln^2\bar x-6x\ln \bar x+\frac{2-x}{3}\ln^3x-\frac{4+3x}{2}\ln^2x
\\ & + 5(x-3)\ln x+10-x\Bigg\}
+C_FT_rN_f\Bigg\{p_{gq}(x)\(\frac{4}{3}\ln^2\bar x+\frac{40}{9}\ln \bar x+\frac{224}{27}\)-\frac{8x}{3}\ln\bar x-\frac{40x}{9}\Bigg\},
\end{align}

\begin{align}
\nn C^{(2,0)}_{g\ot g}(x)&=C_A^2\Bigg\{p_{gg}(x)\Bigg[-24\Li_3(x)+16\ln x\Li_2(x)+4\ln^2 x\,\ln\bar x+4\ln x\,\ln^2\bar x-\frac{2}{3}\ln^3 x+52 \zeta_3
\\\nn &-\frac{808}{27}\Bigg]+p_{gg}(-x)\Bigg[8\Li_3\(\frac{1}{1+x}\)-8 \Li_3\(\frac{x}{1+x}\)+4 \Li_3(x^2)-4\ln x\Li_2(x^2)
\\\nn&+4\ln x\ln^2(1+x)-4\ln x\ln^2(1+x)-\frac{2}{3}\ln^3x -4 \zeta_3\Bigg]+\frac{8}{3}\bar x\(\frac{11}{x}-1+11 x\)\Li_2(\bar x)
\\&\nn -\frac{8}{3}(1+x)\ln^3x
+\frac{44x^2-11x+25}{3}\ln^2x+\frac{2x}{3}\ln \bar x-\frac{536x^2+149 x+701}{9}\ln x
\\ \nn &+\frac{844 x^3-744 x^2+696 x-784}{9x}+\delta(\bar x)\(\frac{1214}{81}-\frac{67\pi^2}{36}-\frac{77}{9}\zeta_3+\frac{5\pi^4}{72}\)
\Bigg\}
\\\nn & + C_A T_rN_f\Bigg\{\frac{224}{27}p_{gg}(x)+\frac{4}{3}(x+1)\ln^2x-\frac{4}{3}x\ln\bar x+\frac{4}{9}(10x+13)\ln x -8\bar x
\\\nn&+ \frac{-332x^3+260}{27 x}+\delta(\bar x)\(-\frac{328}{81}+\frac{5\pi^2}{9}+\frac{28}{9}\zeta_3\)\Bigg\}
\\ &+ C_F T_rN_f\Bigg\{
\frac{4}{3}(1+x)\ln^3x+2(3+x)\ln^2x+24(1+x)\ln x+64 \bar x+\frac{8}{3}\(x^2-\frac{1}{x}\)\Bigg\},
\end{align}

\begin{align}
\nn C^{(2,0)}_{q\ot q'}(x)&=T_r C_F\Bigg\{-\frac{8}{3}\frac{\bar x}{x}(2-x+2x^2)\Li_2(\bar x)+\frac{2}{3}(1+x)\ln^3x-\(1+x+\frac{8x^2}{3}\)\ln^2x
\\& +\frac{4}{9}\(21-30 x+32 x^2\)\ln x+\frac{2}{27}\frac{\bar x}{x}(172-143 x+136 x^2)\Bigg\},
\end{align}

\begin{align}
\nn C^{(2,0)}_{q\ot \bar q}(x)&=\(C_F^2-\frac{C_FC_A}{2}\)\Bigg\{
p_{qq}(-x)\Bigg[8\Li_3\(\frac{1}{1+x}\)-8\Li_3\(\frac{x}{1+x}\)+4\Li_3(x^2)-4\ln x\Li_2(x^2)
\\\nn& +4\ln^2x\ln(1+x)-4\ln x\ln^2(1+x)-\frac{2}{3}\ln^3x-4\zeta_3\Bigg]+4(1+x)\Li_2(1-x^2)-16 \Li_2(\bar x)
\\ & +(22x+6)\ln x+30\bar x
\Bigg\} \,. \label{result:NNLO_pdf_end}
\end{align}

These matching coefficients were first calculated in \cite{Catani:2011kr, Catani:2012qa, Catani:2013tia} by a direct calculation of a cross-section, and in an SCET framework in \cite{Gehrmann:2014yya,Luebbert:2016itl}.
Our results agree with these previous calculations once the proper combination of collinear and soft matrix elements is considered.

%%%%%%%%%%%%%%%%%%%%%%%%%%%%%%%%
\subsection{TMD fragmentation functions}
%%%%%%%%%%%%%%%%%%%%%%%%%%%%%%%%

The LO matching coefficients are
\begin{eqnarray}
\mathbb{C}^{(0,0)}_{q\to q}(z)=\mathbb{C}^{(0,0)}_{g\to g}(z)=\delta(1-z),
\end{eqnarray}
the rest flavor configurations are zero at LO.

The NLO matching coefficients are
\begin{eqnarray}\label{result:NLO_ff_start}
\nn z^2\mathbb{C}^{(1,0)}_{q\to q}(z)&=&C_F\(2\bar z+4 p_{qq}(z)\ln z-\delta(\bar z)\frac{\pi^2}{6}\),\\
\nn z^2\mathbb{C}^{(1,0)}_{q\to g}(z)&=&C_F\(2z+4 p_{gq}(z)\ln z\),\\
\nn z^2\mathbb{C}^{(1,0)}_{g\to q}(z)&=&T_r\(4z\bar z+4p_{qg}(z)\ln z\)\\
\nn z^2\mathbb{C}^{(1,0)}_{g\to g}(z)&=&C_A\(8 p_{gg}(z)\ln z-\frac{\pi^2}{6}\delta(\bar z)\),\\
\mathbb{C}^{(1,0)}_{q\to q'}(z)&=&\mathbb{C}^{(1,0)}_{q\to \bar q}(z)=0.\label{result:NLO_ff_end}
\end{eqnarray}
Here, the coefficient $\mathbb{C}^{(1,0)}_{q\to q'}(z)$ is the coefficient with all the possible mixing flavor channels. Thus $q'$ can be any
quark or anti-quark, even of the same flavor as $q$. In other words, the mathcing coefficient for, say, $u\to u$ is given by the sum
$\mathbb{C}_{q\to q}+\mathbb{C}_{q\to q'}$, as well as the matching coefficient for $u\to\bar u$ is given by $\mathbb{C}_{q\to\bar q}+\mathbb{C}_{q\to q'}$.  The common factor $z^2$ is extracted for convenience. Note that this factor is then not included in the needed
plus-distributions.

The NNLO matching coefficients are
\begin{align}\nn
z^2\mathbb{C}^{(2,0)}_{q\to q}(z)&=C_F^2\Bigg\{p_{qq}(z)\Bigg[40 \Li_3(z)-4\Li_3(\bar z)+4\ln \bar z\Li_2(\bar z)-16\ln z\Li_2(z)-\frac{40}{3}\ln^3 z+ 18 \ln^2 z\ln\bar z
\\ \nn &  -2\ln^2 \bar z \ln z+\frac{15}{2}\ln^2 z-\(8+\frac{4}{3}\pi^2\) \ln z-40 \zeta_3\Bigg]+\bar z\Bigg[24\Li_2(z)+28 \ln z \ln \bar z+10
\\\nn&-\frac{13}{3}\pi^2\Bigg]
+\frac{11}{3}(1+z)\ln^3 z-\frac{59-9z}{2}\ln ^2 z+2\ln\bar z+(46z-38)\ln z+\frac{\pi^4}{72}\delta(\bar z)\Bigg\}
\\\nn & +C_FC_A\Bigg\{p_{qq}(z)\Bigg[
4\Li_3(\bar z)+12\Li_3(z)-4 \ln\bar z\Li_2(\bar z)-8\ln z \Li_2(z)+3\ln^3 z-4 \ln \bar z \ln^2 z
\\\nn &- \frac{11}{6}\ln^2 z+\(\frac{70}{3}-2\pi^2\)\ln z+2 \zeta_3-\frac{404}{27}\Bigg]+4\bar z \Li_2(\bar z)+2(4+z)\ln^2z-2\ln \bar z
\\\nn&  +\frac{116-74 z}{3}\ln z+\frac{44-\pi^2}{3}\bar z+\delta(\bar z)\(\frac{1214}{81}-\frac{67 \pi^2}{36}-\frac{77}{9}\zeta_3+\frac{13\pi^4}{18}\)
\Bigg\}
\\\nn& +C_FT_rN_f\Bigg\{p_{qq}(z)\[\frac{2}{3}\ln^2 z-\frac{20}{3}\ln z+\frac{112}{27}\]-\frac{16}{3}\bar z \ln z-\frac{4}{3}\bar z
\\&+\delta(\bar z)\(-\frac{328}{81}+\frac{5\pi^2}{9}+\frac{28}{9}\zeta_3\)\Bigg\}
 \,,\label{result:NNLO_ff_start}
\end{align}
%%%%%%%%%%%%%%%%%%%
\begin{align}\nn
z^2\mathbb{C}^{(2,0)}_{q\to g}(z)&=
C_F^2\Bigg\{p_{gq}(z)\Bigg[4\Li_3(\bar z)+32 \Li_3(z)+4 \ln\bar z \Li_2(z)-32 \ln z \Li_2(z)-8 \ln \bar z\ln^2 z
\\\nn& +8\ln^2\bar z\ln z-\frac{2}{3}\ln^3 \bar z+\frac{4\pi^2}{3}\ln \bar z+(24-6\pi^2)\ln z-4 \zeta_3\Bigg]+\frac{11}{3}(2-z)\ln^3z
\\\nn& -\(\frac{z}{2}+4\)\ln^2 z+2z\ln^2 \bar z+8 z\ln\bar z\ln z+(25-37 z)\ln z+
2\ln \bar z+(33-3\pi^2)z-38\Bigg\}+
\end{align}
\begin{align}
\\\nn&+C_FC_A\Bigg\{p_{gq}(-z)\Bigg[4 \Li_3\(\frac{1}{1+z}\)-4 \Li_3\(\frac{z}{1+z}\)-2 \Li_3(z^2)-2 \ln z \Li_2(z^2)
\\\nn&-2\ln z \ln^2(1+z)-6\ln^2 z\ln(1+z)+2 \zeta_3\Bigg]+p_{gq}(z)\Bigg[20 \Li_3(z)-4 \Li_3(\bar z)-4 \ln \bar z \Li_2(z)
\\\nn&+\frac{2}{3}\ln^3 \bar z-10\ln^2 \bar z \ln z+22 \ln^2 z\ln \bar z-\frac{4\pi^2}{3}\ln\bar z+\frac{8\pi^2}{3}\ln z-34\zeta_3\Bigg]-32 \ln z \Li_2(z)
\\\nn&-2 z \Li_2(1-z^2)-\frac{8}{3}\(\frac{11}{z}-12+\frac{15z}{2}-2z^2\)\Li_2(\bar z)-\frac{2}{3}\(\frac{40}{z}+22+31 z\)\ln^3 z-2 z \ln^2\bar z
\\\nn&-4 z \ln z\ln\bar z-\frac{4}{3}\(\frac{53}{z}-24+\frac{9z}{4}-6z^2\)\ln^2 z+\frac{2}{3}\(\frac{18}{z}+245+49z+\frac{88z^2}{3}\)\ln z
\\ &-2\ln \bar z+\frac{7\pi^2}{3}z+\frac{4}{3}\(\frac{782}{9 z}-31-\frac{77 z}{2}-\frac{170 z^2}{9}\)\Bigg\},
\end{align}
%%%%%%%%%%%%%%%%%%%%%%%%
\begin{align}\nn
z^2\mathbb{C}^{(2,0)}_{g\to q}(z)&=T_rC_F\Bigg\{p_{qg}(z)\Bigg[32 \Li_3(z)+\frac{2}{3}\ln^3 \bar z-6 \ln^2\bar z\ln z+18 \ln^2 z\ln\bar z+3\ln^2 \bar z-18 \ln \bar z\ln z
\\\nn& +16 \ln \bar z-2\pi^2 \ln \bar z+\frac{2\pi^2}{3}\ln z-3\pi^2 -32\zeta_3\Bigg]+ z\bar z\Bigg[32\Li_2(z)-4\ln^2\bar z+24 \ln z\ln \bar z
\\\nn& -4\ln \bar z-\frac{4\pi^2}{3}\Bigg]-\frac{11}{3}(1-2z+4z^2)\ln^3z-\(\frac{7}{2}+26 z-34 z^2\)\ln^2 z
\\\nn& -(8-73z+76 z^2)\ln z+63-101 z+56 z^2\Bigg\}
\\\nn& +T_rC_A\Bigg\{p_{qg}(-z)\Bigg[4 \Li_3\(\frac{1}{1+z}\)-4 \Li_3 \(\frac{z}{1+z}\)-2 \Li_3(z^2)-2\ln z \Li_2(z^2)
\\\nn&-6 \ln^2 z \ln(1+z)-2 \ln z\ln^2(1+z)+2 \zeta_3\Bigg]+p_{qg}(z)\Bigg[20\Li_3(z)-16 \ln z\Li_2(z)
\\\nn&-\frac{2}{3}\ln^3 \bar z+4\ln^2\bar z\ln z-4\ln^2 z\ln\bar z-\frac{11}{3}\ln^2\bar z+14 \ln z\ln\bar z-\frac{152}{9}\ln \bar z+2\pi^2\ln \bar z-4 \pi^2 \ln z
\\\nn&-6\zeta_3+\frac{19\pi^2}{3}\Bigg]-4z(1+z)\Li_2(1-z^2)+4z\bar z\Bigg[\ln^2 \bar z+\frac{5}{3}\ln \bar z\Bigg]+32z \ln z \Li_2(z)
\\\nn&+\frac{8(2-3z+15z^2-8z^3)}{3z}\Li_2(\bar z)+\frac{2(11+62 z)}{3}\ln^3 z+\frac{2(16-22 z+35 z^2-59z^3)}{3 z}\ln^2 z
\\\nn&+8\ln z\ln\bar z+\frac{2(24-165z-699z^2+38z^3)}{9z}\ln z-\frac{8\pi^2}{3}
\\\nn& -\frac{2(148+1223z-139z^2-774z^3)}{27z}\Bigg\}+T_r^2 N_f\Bigg\{\frac{4}{3}p_{qg}(z)\Bigg[\ln^2 z+\ln^2\bar z-6\ln\bar z\ln z
\\ & -10 \ln z+\frac{10}{3}\ln \bar z-\pi^2+\frac{56}{9}\Bigg]-\frac{16}{3}z\bar z\Bigg[\ln z+\ln \bar z+\frac{2}{3}\Bigg]\Bigg\},
\end{align}
%%%%%%%%%%%%%%%%%%%%%%%%%%%%%%%
\begin{align}\nn
z^2\mathbb{C}^{(2,0)}_{g\to g}(z)&=C_A^2\Bigg\{p_{gg}(-z)\Bigg[8\Li_3\(\frac{1}{1+z}\)-8\Li_3\(\frac{z}{1+z}\)-4\Li_3(z^2)+8\ln z\Li_2(z)
\\\nn &-8\ln z\Li_2(-z)+6 \ln^3z-12\ln^2 z\ln(1+z)-4 \ln z \ln^2(1+z)+4\zeta_3\Bigg]
\\\nn& +p_{gg}(z)\Bigg[104 \Li_3(z)-48 \ln z\Li_2(z)-\frac{62}{3}\ln^3z+28 \ln \bar z\ln^2 z-4\ln^2\bar z\ln z+\frac{44}{3}\ln^2 z
\\\nn&+ \frac{268}{9}\ln z-\frac{20\pi^2}{3}\ln z-\frac{808}{27}-76\zeta_3\Bigg]+\frac{8}{3}\bar z\(1-\frac{11}{z}-11 z\)\Li_2(\bar z)-\frac{88}{3}(1+z)\ln^3 z
\\\nn& + \frac{44z^3-173z^2+103z-264}{3z}\ln^2z+\frac{1340z^3+397z^2+1927z+268}{9z}\ln z-\frac{2}{3}\ln\bar z
\\\nn& + \frac{4(-1064z^3+450z^2-414z+1019)}{27z}+\delta(\bar z)\(\frac{1214}{81}-\frac{67\pi^2}{36}-\frac{77}{9}\zeta_3+\frac{53\pi^4}{72}\)\Bigg\}
\\\nn&+ C_A T_r N_f \Bigg\{p_{gg}(z)\Bigg[-\frac{16}{3}\ln^2 z-\frac{80}{9}\ln z+\frac{224}{27}\Bigg]-\frac{20}{3}(1+z)\ln^2z+\frac{4}{3}\ln \bar z
\\\nn& +\frac{4(26z^3-5z^2+25z-26)}{9z}\ln z+\frac{4(-65z^3+54z^2-54z+83)}{27 z}
\\\nn& +\delta(\bar z)\(-\frac{328}{81}+\frac{5\pi^2}{9}+\frac{28}{9}\zeta_3\)\Bigg\}
\\\nn&+ C_F T_r N_f\Bigg\{\frac{44}{3}(1+z)\ln^3z+\frac{2(16z^3+15z^2+21z+16)}{3z}\ln^2z
\\&-\frac{8(82z^3+81z^2+135 z-6)}{9z}\ln z+\frac{8(301z^3+108z^2-270z-139)}{27 z}\Bigg\},
\end{align}

\begin{align}\nn
z^2\mathbb{C}^{(2,0)}_{q\to q'}(z)&=T_r C_F\Bigg\{\frac{8}{3}\frac{\bar z}{z}(2-z+2z^2)\Li_2(\bar z)+\frac{22}{3}(1+z)\ln^3 z-\(-\frac{32}{3z}+11+11z+8z^2\)\ln^2z
\\&- \frac{4}{9z}(-12+174z+51z^2+32z^3)\ln z-\frac{2}{3}\(\frac{148}{9z}+79-47z-\frac{436}{9}z^2\)
\Bigg\},
\end{align}

\begin{align}\nn
z^2\mathbb{C}^{(2,0)}_{q\to \bar q}(z)&=\(C_F^2-\frac{C_FC_A}{2}\)\Bigg\{p_{qq}(-z)\Bigg[8\Li_3\(\frac{1}{1+z}\)-8\Li_3\(\frac{z}{1+z}\)-4\Li_3(z^2)
\\
\nn& +16\ln z\Li_2(z)-4\ln z \Li_2(z^2)-4 \ln z\ln^2(1+z)-12\ln^2 z\ln(1+z)+6\ln^3z+4\zeta_3\Bigg]
\\& +4(1+z)\Li_2(1-z^2)-16 z \Li_2(\bar z)+8(2+z)\ln^2z+(38-10 z)\ln z+30\bar z
\Bigg\} \,. \label{result:NNLO_ff_end}
\end{align}
The results for the quark sector were first presented by us in \cite{Echevarria:2015usa}\footnote{Concerning the  results for $\mathbb{C}^{(2,0)}_{q\to q}$ we have found a typo in our previous publication \cite{Echevarria:2015usa}.
While we are going to provide a correction for it, we show here the final correct result.}.
The mixed flavor and gluon contributions are presented here for the first time.
Moreover, to the best of our knowledge, the NLO expressions for gluon TMDFF are also presented for the first time.

\section{Matching coefficients at threshold}
\label{sec:conj}

Using our NNLO results for TMDs we observe that it is possible to find a recurrence in the behavior of the matching coefficients for  $x,z\to 1$.

In order to establish the idea, we recall that the matching coefficients for processes where collinear factorization applies, behave as $a_s^k (\ln\,\bar x)^{2k-1}/\bar x$~ \cite{Sterman:1986aj,Catani:1989ne}.  
These corrections are dominant for $x,z\to 1$ and have to be resummed for phenomenological applications (threshold resummation). 
In the case of TMDs, by analyzing the structure of divergences one may expect the leading behavior to be at most like $a_s^k (\ln\,\bar x)^{k-1}/\bar x$. 
In fact, in the case of TMDs the singular behavior at $x,z\to1$ should also be universal, due to the universality of the soft function. 
This statement can be seen in the following way: in the regime $x,z\to1$ the real soft  gluon exchanges in Feynman diagrams are dominant, and are the source of the rapidity divergences in TMD operators. The rapidity divergences are removed by the
$R^f$ factors which are universal for both PDF and FF kinematics. Thus, the leading $x,z\to 1$ behavior of TMDs should be the same. 
At the same time, the leading asymptotic term of the integrated functions is independent of the kinematics and goes like $\sim \Gamma_{cusp}/(1-x)_+$
\cite{Korchemsky:1992xv}, so that we expect the behavior of the matching coefficients to be also universal in the threshold limit.

At two loops, the leading singular behavior at $x,z\to 1$ should be the same for gluons and quarks (up to  a trivial change in the color factor), since it is
produced solely by the convolutions of one-loop soft subgraphs, which are the same for quarks and gluons.   Indeed  for TMDs $F_{f\ot f}$ and $D_{f\to f}$ we observe from our results that
\begin{eqnarray}
F_{f\ot f}^{[2]}&=&C_K^2\(32 \mathbf{L}_\mu^2+\frac{8\pi^2}{3}\)\(\frac{\ln(1-x)}{1-x}\)_++...,
\\\nn
D_{f\to f}^{[2]}&=&C_K^2\(32 \mathbf{L}_\mu^2+\frac{8\pi^2}{3}\)\(\frac{\ln(1-z)}{1-z}\)_++...,
\end{eqnarray}
where dots denote the less dominant contributions and collinear poles. 
The sub-leading contribution, proportional to $1/(1-x)_+$ or $1/(1-z)_+$, is different for gluons and for quarks and depends on $\mathbf{l}_\zeta$, as expected.

We observe that the difference between the gluon and quark channels, as well as the dependence on $\zeta$, disappear after the matching procedure.
In fact, we obtain a simple expression for the leading term at $x,z\to 1$:
\begin{eqnarray}\label{eq:conj}
\nn
C_{f\ot f}^{[2]}&=&16 C_K^2\mathbf{L}_\mu^2\(\frac{\ln(1-x)}{1-x}\)_+
\\ \nn&&-\frac{2C_K}{(1-x)_+}\(2C_K\mathbf{L}_\mu^3
+d^{(2,2)}\mathbf{L}_\mu^2+\(d^{(2,1)}-C_K\frac{\pi^2}{3}\)\mathbf{L}_\mu+d^{(2,0)}\)+...,
\\
\mathbb{C}_{f\to f}^{[2]}&=&16 C_K^2\mathbf{L}_\mu^2\(\frac{\ln(1-z)}{1-z}\)_+
\\\nn&&-\frac{2C_K}{(1-z)_+}\(2C_K\mathbf{L}_\mu^3
+d^{(2,2)}\mathbf{L}_\mu^2+\(d^{(2,1)}-C_K\frac{\pi^2}{3}\)\mathbf{L}_\mu+d^{(2,0)}\)+...,
\end{eqnarray}
where the dots denote the contributions with $\delta$-functions and the non-singular terms at $x,z\to 1$.
The values of $d^{(2,i)}$ can be found in Appendix~\ref{app:ADs}.

In Eq.~(\ref{eq:conj}), the $\m$-scale dependent terms follow from the RGE, while the coefficient for the finite part is peculiar, because it is directly connected to the perturbative expansion of the ${\cal D}^f$ function, which governs the evolution of the TMDs.
If one then extrapolates a similar behavior to an arbitrary loop order, we can make a \textit{conjecture} for the leading term at $x,z,\to 1$ for the finite part of the TMD matching coefficients, based on one- and two-loop calculations:
\begin{eqnarray}\nn
C_{f\ot f}^{(n,0)}&=& \frac{-2C_K}{(1-x)_+}d^{(n,0)}+...~,
\\
\mathbb{C}_{f\to f}^{(n,0)}&=& \frac{-2C_K}{(1-z)_+}d^{(n,0)}+...~.
\end{eqnarray}
The terms proportional to $\mathbf{L}_\mu^k$ can be deduced from the general formulas of Appendix~\ref{app:RGE_rec} in the threshold limit. 
Notice that according to this conjecture, and using the recent result for $d^{(3,0)}$  obtained in~\cite{Li:2016axz}, one can give an estimate of these coefficients at threshold at N$^3$LO.

As a final remark, we notice that, since the soft function enters the polarized TMDs  on the same footing as unpolarized TMDs, a similar result can be obtained for all of them.

%%%%%%%%%%%%%%%%%%%%%%%%%%%%%%%%
%%%%%%%%%%%%%%%%%%%%%%%%%%%%%%%%
%%%%%%%%%%%%%%%%%%%%%%%%%%%%%%%%
\section{Conclusions}
\label{sec:Conclusions}
%%%%%%%%%%%%%%%%%%%%%%%%%%%%%%%%

In this paper we present a comprehensive study of the unpolarized TMDs at NNLO. 
To make it as general as possible, we have introduced the TMD operators, such that TMDs are matrix elements of these operators. 
We find that the understanding of the TMDs benefits from such a language, as provided  by the present work. 
In fact, in these terms, it is possible to introduce a common formalism to describe the universality of soft interactions, the parallelism between the renormalization of UV divergences and rapidity divergences in the TMDs, and their matching onto integrated functions. 
In addition, the consideration of any TMD can be performed without an explicit reference to any given process.

The TMD operators are involved objects, which contain both rapidity divergences as well as UV divergences, and thus are different from usual light-cone operators. 
The rapidity divergences can be absorbed by ``rapidity renormalization factors'', alike the usual UV divergences. 
The explicit form of ``rapidity renormalization factors'' is obtained from the factorization theorems for semi-inclusive DIS, Drell-Yan and $e^+ e^-\to 2 \; hadrons$~\cite{Collins:2011zzd,GarciaEchevarria:2011rb,Echevarria:2012js,Echevarria:2014rua}. 
It is important to note that the ``rapidity renormalization factors'' are the same for all kind of TMD processes, for distribution and fragmentation kinematics and that, together with the UV renomalization, they give direct access to the RGE for TMD operators. That completes the analogy with UV renomalization and it allows to construct a universal TMD operator.

One of the main outcomes of the paper are the matching coefficients of all the unpolarized TMDs onto their integrated analogues. 
According to the operator language they are the Wilson coefficients for the leading term in the small-$b_T$ operator product expansion, as explained in the text. 

The calculation of TMD matrix elements needs a rapidity regulator in addition to a UV regulator. 
For that we have used the (modified) $\delta$-regularization~\cite{Echevarria:2015usa}, in which the form for the ``rapidity renormalization factor'' is especially simple. 
The presented matching coefficients for the TMDPDFs agree with the results of \cite{Gehrmann:2012ze,Gehrmann:2014yya,Luebbert:2016itl}, 
once the proper combination of collinear and soft matrix elements is considered.
Here, instead, we have provided a method that realizes the cancellation of rapidity divergences within a single TMD, and we have checked this fact explicitly at NNLO. 
The results for quark TMDFF were partially presented in~\cite{Echevarria:2015usa}, while here we provide the complete results, which include also the gluon TMDFFs, that were unknown. 
All these matching coefficients are necessary for accurate phenomenological studies, and allow to consider exclusive and inclusive processes on the same level of theoretical accuracy. 

The performed calculation has a complex structure which involves the calculation of TMD matrix elements, integrated matrix elements, TMD soft factor and TMD renormalization constants at NNLO. 
Some of these ingredients are already known at NNLO. So, the TMD soft factor has been presented by our group in \cite{Echevarria:2015byo} and the integrated matrix elements can be related to DGLAP kernels in our regularization scheme. The TMD matrix elements and TMD renormalization constants have been computed at NNLO in this work for the first time with the $\d$ regulator. 
During the calculation of TMD matrix elements we have used many checks, which include: a check of logarithmic parts by RGEs, an independent extraction of anomalous dimensions and a check of the crossing relations between TMDPDF and TMDFF.  
The regularization method described and implemented here, together with the results for the master integrals, can be useful also for the study of polarized TMDs.

In addition, we have studied the limit of large $x$,$z$ and found that the behavior of the matching coefficients is universal for all unpolarized TMDs.
It is naturally controlled by the anomalous dimension ${\cal D}^f$, which allows us to make an all order conjecture on the leading contribution at large $x,z$.

The obtained matching coefficients are necessary in order to pursue phenomenological studies at N$^3$LL accuracy. Some recent developments towards this goal can be found in~\cite{Li:2016ctv}. There, although using a different regulator for rapidity divergences~\cite{Li:2016axz}, the authors have assumed the structure of rapidity divergences which has been explicitly checked in the present work.
Together with the large-$x$ conjecture presented in this work, it opens the door to a very precise estimate of these perturbatively calculable contributions. The phenomenological applications of these results will be exploited in future works. We expect all these efforts to be necessary in order to have a unified picture of Drell-Yan, semi-inclusive DIS and $ e^+e^-\to 2\; hadrons$.

{\bf Note added:} while this article was under submission G.~Lustermans, W.~J.~Waalewijn and L.~Zeune
 \cite{Lustermans:2016nvk}  confirmed  the threshold behavior of the coefficient obtained in Section \ref{sec:conj}.

%%%%%%%%%%%%%%%%%%%%%%%%%%%%%%%%
%%%%%%%%%%%%%%%%%%%%%%%%%%%%%%%%
%%%%%%%%%%%%%%%%%%%%%%%%%%%%%%%%
\acknowledgments
%%%%%%%%%%%%%%%%%%%%%%%%%%%%%%%%

M.G.E. is supported by the Spanish MECD under the \emph{Juan de la Cierva} program and grant FPA2013-46570-C2-1-P.
I.S. is supported by the Spanish MECD grant FPA2014-53375-C2-2-P.

\appendix
%%%%%%%%%%%%%%%%%%%%%%%%%%%%%%%%
%%%%%%%%%%%%%%%%%%%%%%%%%%%%%%%%
%%%%%%%%%%%%%%%%%%%%%%%%%%%%%%%%
\section{NLO expressions}
\label{app:NLO}
%%%%%%%%%%%%%%%%%%%%%%%%%%%%%%%%

For the calculation of the matching coefficients at NNLO, one needs the exact (all-orders in $\epsilon$) expressions for the NLO matching coefficients.
For the details of their calculation see Sec.~\ref{sec:1loop}.
Here we collect all necessary results at NLO.
We use the following notation for some common functions:
\begin{align}\label{def_p}\nn
p_{qq}(x)=\frac{1+x^2}{1-x}\ ,
&&
p_{qg}(x)=1-2x\bar x\ ,
\\
p_{gq}(x)=\frac{1+\bar x^2}{x}~,
&&
p_{gg}(x)=\frac{(1-x\bar x)^2}{x(1-x)}\ .
\end{align}
To denote logarithms throughout the article we use
\begin{eqnarray}\label{app:def_logarithms}
\mathbf{L}_X\equiv \ln\(\frac{X^2\vecb{b}^2_T}{4e^{-2\gamma_E}}\),~~~
\mathbf{l}_X\equiv \ln\(\frac{\mu^2}{X}\),~~~
\pmb \lambda_\delta\equiv\ln\(\frac{\delta^+}{p^+}\).
\end{eqnarray}

The unsubtracted TMDPDFs are
\begin{eqnarray}\nn
\Phi^{[1]}_{q\ot q}&=&2C_F \pmb B^{\epsilon}\Gamma(-\epsilon)\(p_{qq}(x)-\epsilon \bar x-2\delta(\bar x)\pmb \lambda_\delta\),
\\ \nn
\Phi^{[1]}_{q\ot g}&=&2T_r \pmb B^{\epsilon}\Gamma(-\epsilon)\frac{p_{qg}(x)-\epsilon}{1-\epsilon},
\\ \nn
\Phi^{[1]}_{g\ot q}&=&2C_F \pmb B^{\epsilon}\Gamma(-\epsilon)(p_{gq}(x)-\epsilon x),
\\
\Phi^{[1]}_{g\ot g}&=&4C_A \pmb B^{\epsilon}\Gamma(-\epsilon)\(p_{gg}(x)-\delta(\bar x)\pmb \lambda_\delta\) \ ,
\end{eqnarray}
where $\pmb B=\vecb{b}^2_T/4$.
The singularities at $x\to1$ are understood as ``plus''-distribution.
The unsubtracted TMDFFs are
\begin{eqnarray}\nn
z^2\Delta^{[1]}_{q\to q}&=&2C_F \pmb B^{\epsilon}\Gamma(-\epsilon)\(p_{qq}(z)-\epsilon \bar z-2\delta(\bar z)\pmb \lambda_\delta\),
\\ \nn
z^2\Delta^{[1]}_{q\to g}&=& 2C_F 2\pmb B^{\epsilon}\Gamma(-\epsilon)(p_{gq}(z)-\epsilon z),
\\ \nn
z^2\Delta^{[1]}_{g\to q}&=&2T_r \pmb B^{\epsilon}\Gamma(-\epsilon)\frac{p_{qg}(z)-\epsilon}{1-\epsilon},
\\
z^2\Delta^{[1]}_{g\to g}&=&4C_A \pmb B^{\epsilon}\Gamma(-\epsilon)\(p_{gg}(z)-\delta(\bar x)\pmb \lambda_\delta\)\ .
\end{eqnarray}

The matrix elements of the integrated functions are given solely by their UV counterterms. For the PDF kinematics they are
\begin{eqnarray}\nn
f^{[1]}_{q\ot q}&=&\frac{-2C_F}{\epsilon}\(p_{qq}(x)+\frac{3}{2}\delta(\bar x)\),\\ \nn
f^{[1]}_{g\ot q}&=&\frac{-2T_r}{\epsilon}p_{qg}(x),\\ \nn
f^{[1]}_{q\ot g}&=&\frac{-2C_F}{\epsilon}p_{gq}(x),\\
f^{[1]}_{g\ot g}&=&\frac{-1}{\epsilon}\(4C_Ap_{gg}(x)+\(\frac{11}{3}C_A-\frac{4}{3}T_rN_f\)\delta(\bar x)\)\ .
\end{eqnarray}
For the FF kinematics we have
\begin{eqnarray}\nn
d^{[1]}_{q\to q}&=&\frac{-2C_F}{\epsilon}\(p_{qq}(z)+\frac{3}{2}\delta(\bar z)\),\\ \nn
d^{[1]}_{q\to g}&=&\frac{-2C_F}{\epsilon}p_{gq}(z),\\ \nn
d^{[1]}_{g\to q}&=&\frac{-2T_r}{\epsilon}p_{qg}(z),\\
d^{[1]}_{g\to g}&=&\frac{-1}{\epsilon}\(4C_Ap_{gg}(z)+\(\frac{11}{3}C_A-\frac{4}{3}T_rN_f\)\delta(\bar z)\)\ .
\end{eqnarray}

The expression for the NLO soft  factor is
\begin{eqnarray}
S^{[1]}=-4 C_K \pmb B^\epsilon \Gamma(-\epsilon)\(\mathbf{L}_{\sqrt{\zeta}}+2\pmb \lambda_\delta-\psi(-\epsilon)-\gamma_E\),
\end{eqnarray}
where $C_K= C_F (C_A)$ for quark (gluon) case.

For the completeness of exposition we also present the renormalization constants for fields
\begin{align}
Z_2^{[1]}=-\frac{1}{\epsilon}C_F,
&&
Z_3^{[1]}=\frac{1}{\epsilon}\(\frac{5}{3}C_A-\frac{4}{3}T_rN_f\) \ .
\end{align}

%%%%%%%%%%%%%%%%%%%%%%%%%%%%%%%%
%%%%%%%%%%%%%%%%%%%%%%%%%%%%%%%%
%%%%%%%%%%%%%%%%%%%%%%%%%%%%%%%%
\section{NNLO  expressions}
\label{app:NNLO}
%%%%%%%%%%%%%%%%%%%%%%%%%%%%%%%%

In this appendix we present all side expression used for NNLO calculation.

The soft factor at NNLO has been calculated in \cite{Echevarria:2015byo}. 
We present the NNLO contribution to the exponent Eq.~(\ref{eq:S=exp(S1+S2)}). The $\epsilon$-expansion of NNLO soft factor reads
\begin{align}
\label{app:S[2]}\nn
S^{[2]}&=C_K\Bigg[d^{(2,2)}\(\frac{3}{\epsilon^3}+\frac{2\mathbf{l}_\delta}{\epsilon^2}+\frac{\pi^2}{6\epsilon}+\frac{4}{3}\mathbf{L}_\mu^3-2\mathbf{L}_\mu^2\mathbf{l}_\delta
+ \frac{2\pi^2}{3}\mathbf{L}_\mu+\frac{14}{3}\zeta_3\)
\\\nn& -
d^{(2,1)}\(\frac{1}{2\epsilon^2}+\frac{\mathbf{l}_\delta}{\epsilon}-\mathbf{L}_\mu^2+2\mathbf{L}_\mu\mathbf{l}_\delta-\frac{\pi^2}{4}\)
-d^{(2,0)}\(\frac{1}{\epsilon}+2\mathbf{l}_\delta\)
\\&\nn +
C_A\(\frac{\pi^2}{3}+4\,\ln 2\)\(\frac{1}{\epsilon^2}+\frac{2\mathbf{L}_\mu}{\epsilon}
+2\mathbf{L}_\mu^2+\frac{\pi^2}{6}\)+
C_A\(8\,\ln 2-9\zeta_3\)\(\frac{1}{\epsilon}+2\mathbf{L}_\mu\)
+\frac{656}{81}T_RN_f
\\& +C_A\(-\frac{2428}{81}+16\,\ln2-\frac{7\pi^4}{18}-28\,\ln
2\,\zeta_3+\frac{4}{3}\,\pi^2\ln^22-\frac{4}{3}\,\ln^42-32\text{Li}_4\(\frac{1}{2}\)\)+\mathcal{O}(\epsilon)
\Bigg],
\end{align}
where $C_K=C_F(C_A)$ for quark (gluon) soft-factor. Here, the logarithm $\mathbf{l}_\delta$ is $\ln\(\mu^2/|\delta^+\delta^-|\)$, while after substitution Eq.~(\ref{effective_zeta}) it reads
\begin{eqnarray}
\mathbf{l}_\delta=\ln\(\frac{\mu^2}{(\delta^+/p^+)^2\zeta}\)=\mathbf{l}_\zeta-2 \pmb \lambda_\delta.
\end{eqnarray}
The constants $d^{(n,k)}$ are given in Sec.~\ref{app:ADs}.

The NNLO TMD operator constants are calculated in Sec.~\ref{sec:details} and reads
\begin{eqnarray}\nn
Z_q^{[2]}&=&\frac{2C_F^2}{\epsilon^4}+\frac{C_F}{2\epsilon^3}\(8C_F(2+\mathbf{l}_\zeta)+11C_A-4 T_rN_f\)
+\frac{C_F}{\epsilon^2}\Bigg[2C_F(4+4\mathbf{l}_\zeta+\mathbf{l}_\zeta^2)+
\\\nn&&C_A\(\frac{25}{9}+\frac{\pi^2}{6}+\frac{11}{3}\mathbf{l}_\zeta\)-T_rN_f\(\frac{8}{9}+\frac{4}{3}\mathbf{l}_\zeta\)\Bigg]+\frac{C_F}{\epsilon}\Bigg[
C_F\(\pi^2-12\zeta_3\)+
\\&& C_A\(-\frac{355}{27}-\frac{11\pi^2}{12}+13 \zeta_3+\(-\frac{67}{9}+\frac{\pi^2}{3}\)\mathbf{l}_\zeta\)+T_rN_f\(\frac{92}{27}+\frac{\pi^2}{3}+\frac{20}{9}\mathbf{l}_\zeta\)\Bigg],
\end{eqnarray}
\begin{eqnarray}\nn
Z_g^{[2]}&=&\frac{2C_A^2}{\epsilon^4}+\frac{C_A}{2\epsilon^3}\(C_A(19+8\mathbf{l}_\zeta)-4T_rN_f\)+\frac{C_A}{\epsilon^2}\Bigg[
C_A\(\frac{55}{36}+\frac{\pi^2}{6}+\frac{23}{3}\mathbf{l}_\zeta+2\mathbf{l}^2_\zeta\)+
\\&&\nn T_rN_f\(\frac{1}{9}-\frac{4}{3}\mathbf{l}_\zeta\)\Bigg]+\frac{C_A}{\epsilon}\Bigg[C_A\(-\frac{2147}{216}+\frac{11\pi^2}{36}+\zeta_3+\(-\frac{67}{9}+\frac{\pi^2}{3}\)\mathbf{l}_\zeta\)+
\\&& T_rN_f\(\frac{121}{54}-\frac{\pi^2}{9}+\frac{20}{9}\mathbf{l}_\zeta\)\Bigg]~.
\end{eqnarray}
The NNLO field renormalization constants are \cite{Egorian:1978zx}
\begin{eqnarray}\nn
Z_2^{[2]}&=&\frac{C_F}{\epsilon^2}\(\frac{C_F}{2}+C_A\)+\frac{C_F}{\epsilon}\(\frac{3}{4}C_F-\frac{17}{4}C_A+T_rN_f\),
\\
Z_3^{[2]}&=&\frac{C_A}{\epsilon^2}\(-\frac{25}{12}C_A+\frac{5}{3}T_rN_f\)+\frac{1}{\epsilon}\(\frac{23}{8}C_A^2-\frac{5}{2}C_AT_rN_f-2C_F T_rN_f\).
\end{eqnarray}

%%%%%%%%%%%%%%%%%%%%%%%%%%%%%%%%
%%%%%%%%%%%%%%%%%%%%%%%%%%%%%%%%
%%%%%%%%%%%%%%%%%%%%%%%%%%%%%%%%
\section{Results for integrals}
\label{appendix:C}
%%%%%%%%%%%%%%%%%%%%%%%%%%%%%%%%

In this appendix we present the loop integrals that are used to calculate the TMD PDF and TMD FF at NNLO. The parameter $\omega$ is introduced to
in order to resolve the $k^+$ and $l^+$ dependance as explained in Sec.\ref{sec:details}.

%%%%%%%%%%%%%%%%%%%%%%%%%%%%%%%%
\subsection{Integrals for virtual-real diagrams}
%%%%%%%%%%%%%%%%%%%%%%%%%%%%%%%%

The scalar integrals for the virtual-real diagrams has generally the form
\begin{eqnarray}
F^{FF}_{abcde}=-(2\pi)\int\frac{d^dk d^dl}{(2\pi)^{2d}}\frac{p^+\delta(\omega p^++l^+)\delta\(\frac{\bar
z}{z}p^+-k^+\)e^{i(\vecbe{k}\vecbe{b})_T}\delta(k^2)\theta(k^-)}{[(l+p)^2]^a[(k+p)^2]^b[(k+l+p)^2]^c[(k+l)^2]^d[(l^2)]^e}.
\end{eqnarray}
The corresponding integral in PDF kinematics reads
\begin{eqnarray}
F^{PDF}_{abcde}=-(2\pi)\int\frac{d^dk d^dl}{(2\pi)^{2d}}\frac{p^+\delta(\omega p^++l^+)\delta\(\bar x
p^++k^+\)e^{-i(\vecbe{k}\vecbe{b})_T}\delta(k^2)\theta(-k^-)}{[(l+p)^2]^a[(k+p)^2]^b[(k+l+p)^2]^c[(k+l)^2]^d[(l^2)]^e}.
\end{eqnarray}
For our observable only integral with sum of indices equal to 3 contribute.

The momentum $p$ has no transverse component and $p^2=0$. Due to it, the integral with decoupled virtual loop are zero. For example:
$$
F_{02001}=F_{02100}=F_{11001}=F_{11100}=F_{12000}=F_{01110}=F_{02010}=F_{02100}=F_{01011}=0.
$$
Integrals with negative index can be rewritten using identity
$$
(k+l+p)^2+l^2=(p+l)^2+(p+k)^2+(k+l)^2.
$$
The only non-zero integrals with positive indices are
\begin{eqnarray}\nn
F^{FF}_{01101}&=&\frac{-i(-1)^{-\epsilon}}{p^+(4\pi)^d} \frac{\Gamma(-2\epsilon)}{\epsilon}\(\frac{\bar z}{z}\)^\epsilon\pmb
B^{2\epsilon}\frac{z\theta(0<\omega z<1)}{(z\omega(1-z\omega))^\epsilon},
\\
F^{PDF}_{01101}&=&\frac{i}{p^+(4\pi)^d} \frac{\Gamma(-2\epsilon)}{\epsilon}\bar x^\epsilon \pmb
B^{2\epsilon}\frac{\theta(0<\omega/x<1)/x}{(\omega/x(1-\omega/x))^\epsilon},
\end{eqnarray}
\begin{eqnarray}\nn
F^{FF}_{10101}&=&\frac{-i(-1)^{-\epsilon}}{p^+(4\pi)^d} \Gamma(-2\epsilon)\(\frac{\bar z}{z}\)^\epsilon \pmb
B^{2\epsilon}\int[dx]\frac{\delta\(\omega-x_1-\frac{x_2}{z}\)}{(x_2x_3)^{1+\epsilon}},
\\
F^{PDF}_{10101}&=&\frac{i}{p^+(4\pi)^d} \Gamma(-2\epsilon)\bar x^\epsilon\pmb B^{2\epsilon}\int[dx]\frac{\delta\(\omega-x_1-x
x_2\)}{(x_2x_3)^{1+\epsilon}},
\end{eqnarray}
where $[dx]=\delta(1-x_1-x_2-x_3)dx_1dx_2dx_3$. We leave the integral over Feynman parameters in  $F_{10101}$, since it is convenient first over
$\omega$ with the help of $\delta$-function. There are two another integrals that appear in calculation and can be reduced to the previous cases
\begin{eqnarray}\nn
F_{00111}\(\omega\)&=&F_{10101}\(\frac{p^++k^+}{p^+}-\omega\),
\\ \nn
F^{FF}_{021(-1)1}&=&-\(z\omega+\bar z\(1-2z\omega\)\)F^{FF}_{01101},
\\
F^{PDF}_{021(-1)1}&=&-\(\frac{\omega}{x}-\frac{\bar x}{x}\(1-2\frac{\omega}{x}\)\)F^{PDF}_{01101}.
\end{eqnarray}

%%%%%%%%%%%%%%%%%%%%%%%%%%%%%%%%
\subsection{Integrals for double-real diagrams}
%%%%%%%%%%%%%%%%%%%%%%%%%%%%%%%%

The scalar integrals for the double-real diagrams have generally the form
\begin{eqnarray}
F_{abcd}=(2\pi)^2\int\frac{d^{d-1}k
d^{d-1}l}{(2\pi)^{2d}}\frac{e^{i(\vecbe{k}\vecbe{b})_T}\,
e^{i(\vecbe l\vecbe k)_T}
\delta(k^2)\theta(k^-)\delta(l^2)\theta(l^-)}{[(l+p)^2]^a[(k+p)^2]^b[(k+l+p)^2]^c[(k+l)^2]^d}.
\end{eqnarray}
The components $k^+$ and $l^+$ must be integrated with the help of $\delta$-functions as explained in Sec.\ref{sec:details} and do not
participate in the loop-integration (that is indicated by $d-1$-dimensional integral). Integrating over minus components using on-mass-shell
$\delta$-functions we arrive to standard euclidian loop integral over transverse momentum. The theta function on the minus-components implies the $k^+,l^+>0$ in the result of integration.

In our calculation only the integral with sum of indices equal to $2$ participate. Here is the list of non-zero integrals
\begin{eqnarray}\nn
F_{0110}&=&\frac{-1}{(4\pi)^{d}}\pmb
B^{2\epsilon}\frac{\Gamma(-2\epsilon)}{\epsilon}
\\&&\nn\frac{1}{k^++p^+}\(\frac{l^+p^+(k^++p^++l^+)}{k^+(k^++p^+)^2}\)^{\epsilon}
~_2F_1\(-\epsilon,-2\epsilon,1-\epsilon;\frac{-k^+(k^++p^++l^+)}{p^+l^+}\),
\\ \nn
F_{1010}&=&\frac{-1}{(4\pi)^{d}}\pmb
B^{2\epsilon}\frac{\Gamma(-2\epsilon)}{\epsilon}\\&&\nn\frac{1}{l^++p^+}\(\frac{k^+p^+(k^++p^++l^+)}{l^+(l^++p^+)^2}\)^{\epsilon}
~_2F_1\(-\epsilon,-2\epsilon,1-\epsilon;\frac{-l^+(k^++p^++l^+)}{p^+k^+}\),
\\ \nn
F_{1100}&=&\frac{1}{(4\pi)^{d}}\pmb B^{2\epsilon}\frac{\Gamma^2(-\epsilon)}{p^+},
\\ \nn
F_{0020}&=&\frac{1}{(4\pi)^{d}}\pmb B^{2\epsilon}\frac{\Gamma(-2\epsilon)}{k^++l^++p^+}\(\frac{(k^++l^+)^2(k^++l^++p^+)}{k^+l^+p^+}\)^\epsilon,
\\ \nn
F_{1001}&=&\frac{1}{(4\pi)^{d}}\pmb B^{2\epsilon}\frac{\Gamma^2(-\epsilon)}{l^+}\(\frac{k^++l^+}{l^+}\)^{2\epsilon},
\\
F_{0101}&=&\frac{1}{(4\pi)^{d}}\pmb B^{2\epsilon}\frac{\Gamma^2(-\epsilon)}{k^+}\(\frac{k^++l^+}{k^+}\)^{2\epsilon},
\end{eqnarray}
where $\pmb B=\vecb{b}_T^2/4$. The integrals with negative indices can be obtained from the ones presented here, by differentiation with respect to $k^+$
or $l^+$.

%%%%%%%%%%%%%%%%%%%%%%%%%%%%%%%%
%%%%%%%%%%%%%%%%%%%%%%%%%%%%%%%%
%%%%%%%%%%%%%%%%%%%%%%%%%%%%%%%%
\section{Recursive relations from RGE and anomalous dimensions}
\label{appendix:D}
%%%%%%%%%%%%%%%%%%%%%%%%%%%%%%%%

In this Appendix we collect all the expressions necessary for the application of the RGEs, as well as the explicit expressions for the logarithmical part of the matching coefficients.

%%%%%%%%%%%%%%%%%%%%%%%%%%%%%%%%
\subsection{Recursive form of RGEs}
\label{app:RGE_rec}
%%%%%%%%%%%%%%%%%%%%%%%%%%%%%%%%

The derivation of RGEs is given in Section \ref{sec:RGE}.
The $\zeta$-dependence of the matching coefficients can be explicitly solved by Eq.~(\ref{RGE:zeta_dep}).
The $\mu$-dependence is then given by the RGEs in Eq.~(\ref{RGE:C_mu}).
For practical purposes it is convenient to rewrite the RGE application in a recursive form. We use the notation of Eq.~(\ref{eq:Clognotation}).
%\begin{eqnarray}
%C^{[n]}_{f\ot f'}(x,\mathbf{L}_\mu)&=&\sum_{k=0}^{2n}C_{f\ot f'}^{(n;k)}(x)\mathbf{L}_\mu^k, \nn
%\\
%\mathbb{C}^{[n]}_{f\to f'}(x,\mathbf{L}_\mu)&=&\sum_{k=0}^{2n}\mathbb{C}_{f\to f'}^{(n;k)}(z)\mathbf{L}_\mu^k.
%\end{eqnarray}
Then the equation of the logarithmic dependent part of the TMDPDF matching coefficient reads
\begin{align}\label{app:rec_RGE1}%\nn
(k+1)C_{f\ot f'}^{(n;k+1)}&=\sum_{r=1}^n\Big[\frac{\Gamma^{f(r)}}{2}C_{f\ot f'}^{(n-r;k-1)}
%\\&%&\nn \quad\quad
+\((n-r)\beta^{(r)}-\frac{\gamma^{f(r)}_V}{2}\)C_{f\ot f'}^{(n-r;k)}-C_{f\ot h}^{(n-r;k)}\otimes P^{(r)}_{h\ot f'}(x)\Big]
\,.
\end{align}
The same logarithmic part of  the TMDFF matching coefficient reads
\begin{align}\label{app:rec_RGE2}\nn
(k+1)\mathbb{C}_{f\to f'}^{(n;k+1)}&=\sum_{r=1}^n\Big[\frac{\Gamma^{f(r)}}{2}\mathbb{C}_{f\to f'}^{(n-r;k-1)}
\\&% \nn \quad\quad
+\((n-r)\beta^{(r)}-\frac{\gamma^{f(r)}_V}{2}\)\mathbb{C}_{f\to f'}^{(n-r;k)}
-\mathbb{C}_{f\to h}^{(n-r;m)}\otimes \(\frac{\mathbb{P}^{(r)}_{h\to f'}(z)}{z^2}\)\Big].
\end{align}
Solving Eq.~(\ref{app:rec_RGE1}-\ref{app:rec_RGE2}) at NLO we obtain
\begin{align}
C^{(1;2)}_{f\ot f'}=\delta_{ff'}\delta(\bar x)\frac{\Gamma^{f(1)}}{4},
&&
C^{(1;1)}_{f\ot f'}=-\delta_{ff'}\delta(\bar x)\frac{\gamma_V^{f(1)}}{2}-P^{(1)}_{f\ot f'}(x)
\,.
\end{align}
At NNLO we finally have
\begin{eqnarray}\nn
C^{(2;4)}_{f\ot f'}&=&\delta_{ff'}\delta(\bar x)\frac{\(\Gamma^{f(1)}\)^2}{32},
\\ \nn
C^{(2;3)}_{f\ot f'}&=&\delta_{ff'}\delta(\bar
x)\frac{\Gamma^{f(1)}}{4}\(\frac{\beta^{(1)}}{3}-\frac{\gamma_V^{f(1)}}{2}\)-\frac{\Gamma^{f(1)}P^{(1)}_{f\ot f'}(x)}{4},
\\ \nn
C^{(2;2)}_{f\ot f'}&=&\delta_{ff'}\delta(\bar x)\(\frac{\Gamma^{f(2)}}{4}-\frac{\gamma_V^{f(1)}\beta^{(1)}}{4}+\frac{(\gamma_V^{f(1)})^2}{8}\)+
\\\nn &&P_{f\ot f'}^{(1)}\frac{\gamma_V^{f(1)}-\beta^{(1)}}{2}+C^{(1;0)}_{f\ot f'}\frac{\Gamma^{f(1)}}{4}+\frac{1}{2}\sum_rP^{(1)}_{f\ot r}\otimes P^{(1)}_{r\ot f'},
\\
C^{(2;1)}_{f\ot f'}&=&-\delta_{ff'}\delta(\bar x)\frac{\gamma_V^{f(2)}}{2}-P_{f\ot f'}^{(2)}+C^{(1;0)}_{f\to
f'}\(\beta^{(1)}-\frac{\gamma_V^{f(1)}}{2}\) -\sum_r C^{(1;0)}_{f\ot r}\otimes P^{(1)}_{r\ot f'}.
\end{eqnarray}
The expressions for TMDFF matching coefficients can be obtained from these ones by changing the directions of the arrows and replacing DGLAP kernels as $P\to \mathbb{P}/z^2$.
Explicit expression for these equations can be found in a supplementary file~\cite{file}.

%%%%%%%%%%%%%%%%%%%%%%%%%%%%%%%%
\subsection{Anomalous dimensions}
\label{app:ADs}
%%%%%%%%%%%%%%%%%%%%%%%%%%%%%%%%

For the calculation at NNLO one needs the following anomalous dimensions:
\begin{itemize}
\item the QCD $\beta$-function, $\b(\as)=d\as/d\ln\m$, with $\b= -2\as \sum_{n=1}^\infty \b^{(n)} \left( \frac{\as}{4\pi} \right)^n$
\begin{align}
\beta^{(1)}&=\frac{11}{3}C_A-\frac{4}{3}T_rN_f\equiv b_0\,,
\nn
\\
\b^{(2)} &=
\frac{34}{3}\,C_A^2 - \frac{20}{3}\,C_A T_r N_f
- 4 C_F T_r N_f
%\approx 38.6667
\,,
\nn\\
\b^{(3)}  &=
\frac{2857}{54}\,C_A^3 + \left( 2 C_F^2
- \frac{205}{9}\,C_F C_A - \frac{1415}{27}\,C_A^2 \right) T_r N_f
%\nn\\
%&
+ \left( \frac{44}{9}\,C_F + \frac{158}{27}\,C_A \right) T_r^2 N_f^2
%\nn\\
%& \approx
%180.907
\,,
\nn\\
\b^{(4)}  &=
\frac{149753}{6} + 3564\zeta_3
- \left( \frac{1078361}{162} + \frac{6508}{27}\,\zeta_3 \right) N_f
%\nn\\
%&
+ \left( \frac{50065}{162} + \frac{6472}{81}\,\zeta_3 \right) N_f^2
+ \frac{1093}{729}\,N_f^3
%\nn\\
%&\approx
%4826.16
\,,
\end{align}
\item the cusp anomalous dimension
\begin{align}
\Gamma_{cusp}^{q}&=4C_F \Gamma,~~~~\Gamma^g_{cusp}=4C_A\Gamma,
\nn \\
\Gamma^{(1)}&=1,\qquad \Gamma^{(2)}=\(\frac{67}{9}-\frac{\pi^2}{3}\)C_A-\frac{20}{9}T_rN_f.\nn
\\
\G^{(3)} &=
 C_A^2 \left( \frac{245}{6} - \frac{134\pi^2}{27}
+ \frac{11\pi^4}{45} + \frac{22}{3}\,\zeta_3 \right)
+ C_A T_r N_f  \left( - \frac{418}{27} + \frac{40\pi^2}{27}
- \frac{56}{3}\,\zeta_3 \right)
\nn\\
&
+ C_F T_r N_f \left( - \frac{55}{3} + 16\zeta_3 \right)
- \frac{16}{27}\,T_r^2 N_f^2
\end{align}
\item the anomalous dimension $\gamma_V$
\begin{align}
\gamma_V^{q(1)}&=-6C_F,\nn
\\ \nn \gamma^{q(2)}_V&=C_F^2\(-3+4\pi^2-48\zeta_3\)+C_FC_A\(-\frac{961}{27}-\frac{11\pi^2}{3}+52\zeta_3\)+C_FT_rN_f\(\frac{260}{27}+\frac{4\pi^2}{3}\),
\\
\gamma_V^{q(3)}&=
C_F^3 \left( -29 - 6\pi^2 - \frac{16\pi^4}{5}
- 136\zeta_3 + \frac{32\pi^2}{3}\,\zeta_3 + 480\zeta_5 \right)
\nn\\
&
+ C_F^2 C_A \left( - \frac{151}{2} + \frac{410\pi^2}{9}
+ \frac{494\pi^4}{135} - \frac{1688}{3}\,\zeta_3
- \frac{16\pi^2}{3}\,\zeta_3 - 240\zeta_5 \right)
\nn\\
&
+ C_F C_A^2 \left( - \frac{139345}{1458} - \frac{7163\pi^2}{243}
- \frac{83\pi^4}{45} + \frac{7052}{9}\,\zeta_3
- \frac{88\pi^2}{9}\,\zeta_3 - 272\zeta_5 \right)
\nn\\
&
+ C_F^2 T_r N_f \left( \frac{5906}{27} - \frac{52\pi^2}{9}
- \frac{56\pi^4}{27} + \frac{1024}{9}\,\zeta_3 \right)
\nn\\
&
+ C_F C_A T_r N_f \left( - \frac{34636}{729}
+ \frac{5188\pi^2}{243} + \frac{44\pi^4}{45} - \frac{3856}{27}\,\zeta_3
\right)
+ C_F T_r^2 N_f^2 \left( \frac{19336}{729} - \frac{80\pi^2}{27}
- \frac{64}{27}\,\zeta_3 \right)
\,,
\end{align}
%%%%%%%%%
\begin{align}
 \gamma^{g(1)}_V&=-\frac{22}{3}C_A+\frac{8}{3}T_rN_f,
\nn\\
\gamma_V^{g(2)}&=C_A^2\(-\frac{1384}{27}+\frac{11\pi^2}{9}+4\zeta_3\)+C_AT_rN_f\(\frac{512}{27}-\frac{4\pi^2}{9}\)+8C_FT_rN_f.
\nn\\
\gamma_V^{g(3)}&=2 C_A^3\left(\frac{-97186}{729}+\frac{6109}{486}\pi^2-\frac{319}{270}\pi^4+\frac{122}{3}\zeta_3-\frac{20}{9}\pi^2\zeta_3-16\zeta_5\right)\nn \\
&+2 C_A^2 T_r N_f\left(\frac{30715}{729}-\frac{1198}{243}\pi^2+\frac{82}{135}\pi^4+\frac{712}{27}\zeta_3\right)\nn \\
&+2 C_A C_F T_r N_f\left(\frac{2434}{27}-\frac{2}{3}\pi^2-\frac{8}{45}\pi^4-\frac{304}{9}\zeta_3\right)
-4 C_F^2 T_r N_f\nn \\
&+
2 C_A T_r^2 N_f^2 \left(-\frac{538}{729}+\frac{40}{81}\pi^2-\frac{224}{27}\zeta_3\right)
-\frac{88}{9} C_F T_r^2 N_f^2
\end{align}
\item It is convenient to write the expression for the function $\mathcal{D}$ as an expansion:
\begin{eqnarray}
\mathcal{D}^{f}(\mu,\vecb{b}_T)&=&C^f\sum_{n=1}^\infty a_s^n \sum_{k=0}^n \mathbf{L}^k_\mu d^{(n,k)},
\end{eqnarray}
where $C^f=C_F$ for quarks and $C^f=C_A$ for gluons, and
\begin{eqnarray*}
d^{(1,1)}&=&2\Gamma^{(1)},\qquad d^{(1,0)}=0,
\\
d^{(2,2)}&=&\Gamma^{(1)}\beta^{(1)},\qquad d^{(2,1)}=2\Gamma^{(2)},
\\
d^{(2,0)}&=&C_A\(\frac{404}{27}-14\zeta_3\)-\frac{112}{27}T_rN_f.
\\
d^{(3,3)}&=&\frac{2}{3}\Gamma^{(1)}(\beta^{(1)})^2,
\qquad d^{(3,2)}=2 \Gamma^{(2)}\beta^{(1)}+ \Gamma^{(1)}\beta^{(2)}\ , \\
d^{(3,1)}&=& 2 \beta^{(1)} d^{(2,0)}+2\Gamma^{(3)}, \\
d^{(3,0)}&=& \frac{-1}{2}C_A^2 \left(-\frac{176}{3}\zeta_3\zeta_2+\frac{6392\zeta_2}{81}+\frac{12328\zeta_3}{27}+
\frac{154\zeta_4}{3}-192 \zeta_5-\frac{297029}{729}\right)
\nn \\ &&
- C_A T_r N_f\left(-\frac{824\zeta_2}{81}-\frac{904\zeta_3}{27}+\frac{20\zeta_4}{3}+\frac{62626}{729}\right)
- 2 T_r^2 N_f^2 \left(-\frac{32\zeta_3}{9}-\frac{1856}{729}\right)
\nn \\ &&
- C_F T_r N_f \left(\frac{-304\zeta_3}{9}-16\zeta_4+\frac{1711}{27}\right)\, .
\end{eqnarray*}
The result for  $d^{(3,0)}$ has been recently computed in \cite{Li:2016ctv}. The rest of $d^{(3,i)}$  can be found also in \cite{Echevarria:2012pw}.
\end{itemize}

The DGLAP kernels at LO read
\begin{eqnarray}\nn
P^{(1)}_{q\ot q}(x)&=&C_F\(2p_{qq}(x)+3\delta(\bar x)\),\qquad\qquad\mathbb{P}^{(1)}_{q\to q}(z)=C_F\(2p_{qq}(z)+3\delta(\bar z)\),
\\ \nn
P^{(1)}_{q\ot g}(x)&=&2T_rp_{qg}(x),\qquad\qquad\qquad\qquad~~~ \mathbb{P}^{(1)}_{g\to q}(z)=2C_Fp_{qg}(z),
\\ \nn
P^{(1)}_{g\ot q}(x)&=&2C_Fp_{gq}(x),\qquad\qquad\qquad\qquad~~ \mathbb{P}^{(1)}_{q\to g}(z)=2T_rp_{qg}(z),
\\
P^{(1)}_{g\ot g}(x)&=&4C_Ap_{gg}(x)+\beta^{(1)}\delta(\bar x),\qquad\qquad \mathbb{P}^{(1)}_{g\to
g}(z)=4C_Ap_{gg}(z)+\beta^{(1)}\delta(\bar z).
\end{eqnarray}
The NLO kernels for PDF kinematic can be found in \cite{Moch:1999eb}, for FF kinematic in \cite{Mitov:2006wy}.

%%%%%%%%%%%%%%%%%%%%%%%%%%%%%%%%
%%%%%%%%%%%%%%%%%%%%%%%%%%%%%%%%
%%%%%%%%%%%%%%%%%%%%%%%%%%%%%%%%
\section{Alternative form of matching coefficients}
%%%%%%%%%%%%%%%%%%%%%%%%%%%%%%%%

For practical purposes, it is convenient to write the matching coefficients as overall ``plus''-distributions.
In this Appendix we rewrite the expressions for the matching coefficients in such a form.
Only the flavor-diagonal coefficients need to be rewritten in this way, since the non-diagonal channels are integrable at $z,x\to 1$.

The NLO expressions read
\begin{eqnarray}\nn
C_{q\ot q}^{(1,0)}(x)&=&\(C_{q\ot q}^{(1,0)}(x)\)_++\delta(\bar x)C_F\(1-\frac{\pi^2}{6}\),
\\ \nn
C_{g\ot g}^{(1,0)}(x)&=&\frac{1}{x}\(xC_{g\ot g}^{(1,0)}(x)\)_+-\delta(\bar x)C_A\frac{\pi^2}{6},
%\end{eqnarray}
%\begin{eqnarray}
\\ \nn
\mathbb{C}_{q\to q}^{(1,0)}(z)&=&\frac{1}{z^2}\(z^2\mathbb{C}_{q\to q}^{(1,0)}(z)\)_++\delta(\bar z)C_F\(6-\frac{3}{2}\pi^2\),
\\
\mathbb{C}_{g\to g}^{(1,0)}(z)&=&\frac{1}{z^3}\(z^3\mathbb{C}_{g\to g}^{(1,0)}(z)\)_++\delta(\bar z)C_A\(\frac{65}{18}-\frac{3}{2}\pi^2\),
\end{eqnarray}
where the matching coefficients for TMDPDF and TMDFF case on the r.h.s. are taken from Eq.~(\ref{result:NLO_pdf_start}) and
Eq.~(\ref{result:NLO_ff_start}) respectively.
Obviously, only regular parts of the matching coefficients on the r.h.s. contribute, since $(\delta(\bar z))_+=0$.

The NNLO expressions are
\begin{eqnarray}\nn
C_{q\ot q}^{(2,0)}(x)&=&\(C_{q\ot q}^{(2,0)}(x)\)_++\delta(\bar x)C_F\Bigg[C_F\(\frac{203}{8}-\frac{25\pi^2}{6}-12\zeta_3+\frac{157\pi^4}{360}\)
\\&&+C_A\(\frac{7277}{324}+\frac{175\pi^2}{108}-\frac{278}{9}\zeta_3-\frac{7\pi^4}{30}\)+T_rN_f\(-\frac{1565}{162}-\frac{5\pi^2}{27}+\frac{52}{9}\zeta_3\)\Bigg],
\\ \nn
C_{g\ot g}^{(2,0)}(x)&=&\frac{1}{x}\(xC_{g\ot g}^{(2,0)}(x)\)_++\delta(\bar x)\Bigg[C_A^2\(\frac{16855}{324}-\frac{113\pi^2}{36}-\frac{407}{9}\zeta_3+\frac{53\pi^4}{360}\)
\\
&&+C_A T_rN_f\(-\frac{577}{81}+\frac{5\pi^2}{9}+\frac{28}{9}\zeta_3\)+C_FT_rN_f\frac{85}{81}\Bigg],
\end{eqnarray}
\begin{eqnarray}
\mathbb{C}_{q\to q}^{(2,0)}(z)&=&\frac{1}{z^2}\( z^2\mathbb{C}_{q\to q}^{(2,0)}(z)\)_++\delta(\bar z)C_F\Bigg[C_F\(-\frac{213}{8}-5\pi^2-12\zeta_3+\frac{397\pi^4}{360}\)
\\\nn&&+C_A\(\frac{6353}{81}-\frac{443\pi^2}{36}-\frac{278}{9}\zeta_3+\frac{91\pi^4}{90}\)+T_rN_f\(-\frac{2717}{162}+\frac{25\pi^2}{9}+\frac{52}{9}\zeta_3\)\Bigg],
\\
\mathbb{C}_{g\to g}^{(2,0)}(z)&=&\frac{1}{z^3}\(z^3\mathbb{C}_{g\to g}^{(2,0)}(z)\)_++\delta(\bar z)\Bigg[C^2_A\(43-\frac{430\pi^2}{27}-\frac{605}{9}\zeta_3+\frac{59\pi^4}{24}\)
\\\nn&&+C_A T_r N_f\(\frac{38}{81}+\frac{55\pi^2}{27}-\frac{68}{9}\zeta_3\)+C_FT_rN_f \frac{674}{81}\Bigg].
\end{eqnarray}


\end{document}